\documentclass[pre,aps,onecolumn,superscriptaddress]{revtex4}
\usepackage{graphicx}

\begin{document}

\title{
Emergence of Unstable Modes in an Expanding Domain  \\ 
for
Energy-Conserving Wave Equations}
\author{K.J.H. Law$^1$, P.G. Kevrekidis}
\affiliation{Department of Mathematics and Statistics, University of
Massachusetts, Amherst MA 01003-4515}
\author{D.J. Frantzeskakis}
\affiliation{Department of Physics, University of Athens, Panepistimiopolis,
Zografos, Athens 15784, Greece}
\author{A.R. Bishop}
\affiliation{Theoretical Division and Center for
Nonlinear Studies, Los
Alamos National Laboratory, Los Alamos, New Mexico 87545, USA}

\begin{abstract}
Motivated by recent work on instabilities in expanding domains in
reaction-diffusion settings, we propose an analog of such mechanisms
in energy-conserving
wave equations.
In particular, we consider
a nonlinear Schr{\"o}dinger equation in a finite domain and show how
the expansion or contraction of the domain, under appropriate conditions,
can destabilize its originally stable solutions through the 
modulational instability mechanism. Using both real and Fourier space
diagnostics, we
monitor and control the crossing of the instability
threshold and, hence, the activation of the instability. We also
consider how the manifestation of this mechanism is modified
in a spatially inhomogeneous
setting, namely in the presence of an external parabolic potential,
which is relevant to trapped Bose-Einstein condensates.
\end{abstract}

\maketitle

\section{Introduction}

There has been a considerable recent interest in the
study of pattern formation in expanding domains, and pattern formation
supported by  them,
in the context of reaction-diffusion equations (see, e.g.,
\cite{crampin1,crampin2,crampin3,maini}
and references therein for a substantial body of earlier work).
Such mechanisms are viewed in mathematical biology
as a possibly important element in the
formation of skin patterns on species of
animals or fishes \cite{murray}. One of the key ideas in that
regard is that in the case of finite-size domains, Turing-type mechanisms \cite{turing}
are only activated for appropriate wavenumbers, which, in turn, signify
appropriate sizes of the domain. In particular, as is pointed out
in \cite{crampin1}, the domain size should be such so as to admit at
least a half-cycle of the intrinsic pattern (to be formed by the instability) wavelength.
Once activated, these mechanisms typically lead to the
destabilization of a uniform concentration profile, due to the
presence of diffusion, and the formation of localized waveforms
through frequency-doubling and related mechanisms~\cite{crampin1,crampinphd}.

Our aim in the present work is to illustrate that such mechanisms
are not only relevant in reaction-diffusion equations and pattern-forming
dissipative systems, but also to Hamiltonian 
nonlinear dispersive wave equations, such as the prototypical model of the nonlinear
Schr{\"o}dinger (NLS) equation \cite{sulem}. The latter class of models
also features instabilities which
emerge only for appropriate bands of wavenumbers;
a prominent example of such instabilities
is the
modulational instability (MI) of the plane wave solutions of the focusing NLS equation
(see e.g. \cite{kf} for a recent review and \cite{hasegawa} for a detailed discussion).
This instability only emerges (for a fixed amplitude of the plane wave)
for sufficiently {\it small} wavenumbers, i.e., for sufficiently large perturbation wavelengths,
or (for a
fixed wavenumber) for sufficiently large amplitudes
of the
plane wave. Hence, one can
envision a setting where the initial
domain is small enough that the smallest wavenumber permissible
by the domain (say, $2 \pi/L$ for periodic boundary conditions and
domain size $L$) is larger than the critical wavenumber for the onset of the instability.
Equivalently, the domain size may be smaller, at time $t=0$, than the
minimum wavelength necessary for the instability. Then, 
with an appropriate temporal dependence, the domain size may increase 
and cross the relevant instability threshold; this, in turn, leads to the emergence of patterns 
in a way reminiscent of that 
occuring in their Turing analogs in 
dissipative systems. In fact, these patterns in the Hamiltonian case will consist of robust 
solitary waves, as we will illustrate below, since MI is one of the key mechanisms leading to the formation of such 
structures \cite{kf,hasegawa}.

This mechanism could be of relevance to a wide array of
Hamiltonian systems, since the NLS model is relevant to a diverse
set of applications
including the propagation of light in optical fibers \cite{hasegawa},
the dynamics of Bose-Einstein condensates (BECs) at extremely low temperatures \cite{BEC},
plasma physics and fluid dynamics \cite{infeld}, and so on.
In fact, 
MI 
in itself is still one of the most active topics of investigation in some of these
contexts. In particular, a variant of MI was used in the first
experiment that demonstrated the generation of 
matter-wave soliton trains in BECs \cite{rice}. Furthermore, MI is recently becoming
increasingly accessible in a variety of settings experimentally
(and theoretically); these include spatially periodic ones 
such as BECs confined in optical lattices (where MI introduces a classical
form of a superfluid-insulator transition \cite{augusto}) and
optical waveguide arrays \cite{dnc0}, as well as ones that are
periodic in the evolution variable, such as the nonlinear
layered optical media recently studied in Refs.  \cite{mason}.

The presentation of our results will be structured as follows. We
will first give a brief theoretical background on the nature of
the instability and explain how it is affected by the existence of
a finite domain size. We will then proceed to examine four
different numerical experiments, attempting to elucidate the emergence
of the instability for time-dependent domain sizes. Our first and
most straightforward example will be the one discussed above:
a stable uniform solution (for sufficiently small domain size) will
be subjected to a time-dependent domain expansion, resulting in what
we will call a ``soliton sprinkler'' through the emergence of MI.
Our second experiment will be somewhat counter-intuitive
in that the instability will emerge even though the domain size is
{\it decreasing}. This will be achieved by preserving the $L^2$
norm of the solution (its power in optics or its number of atoms in BEC),
and taking advantage of the instability dependence on the beam intensity.
The last two examples will introduce an expanding or
contracting domain situation in a somewhat modified but also
directly relevant and more straightforwardly realizable experimental
setting, inspired by the physics of BECs.
In particular, we will use a parabolic potential (routinely employed to 
trap the BEC) with a time-dependent
frequency, which, in turn, may weaken or strengthen the confinement of
the atoms, introducing a corresponding {\it effective} expansion or
contraction of the domain. We will seek signatures of MI in this
more complex, yet more realistic setting. Finally, we will
summarize our findings and present our conclusions.

\section{Results}

\subsection{Theoretical Background}

We consider the following dimensionless NLS equation for the field $u(x,t)$,
%
\begin{equation}
i \partial_t u =
-\partial_{x}^{2} u + g|u|^2 u + V_{\rm ext}(x)u,
\label{gp}
\end{equation}
where $g$ is the nonlinearity coefficient and $V_{\rm ext}(x)$ is an external potential.
Note that in the context of BECs this equation is usually referred to as the Gross-Pitaevskii
(GP) equation \cite{BEC}); in that case, $u(x,t)$ is the BEC wavefunction, $g$ is
proportional to the s-wave scattering length, and $V_{\rm ext}(x)$ is a trapping
potential where the BEC is confined. For simplicity, the nonlinearity coefficient is
scaled so that $g = \pm 1$, corresponding, respectively, to repulsive or attractive interatomic interactions
(or, generally, to defocusing or focusing nonlinearity). Equation (\ref{gp})
conserves the energy, as well as the squared $L^2$ norm, $N = \int_{-\infty}^{+\infty} |u|^2 dx$ (in the context
of BECs, $N$ represents the normalized number of atoms of the condensate).

For our theoretical analysis, we first consider the homogeneous (untrapped)
case, i.e., Eq. (\ref{gp})
with $V_{\rm ext}=0$, which admits plane-wave solutions of the form
$u(x,t)=u_0 \exp[i(kx-\omega t)]$, where $u_{0}$, $k$, and
$\omega$ are, respectively, the amplitude, wavenumber and frequency of the
plane wave 
solution. These parameters are connected
through the dispersion relation: $\omega=k^2+g u_0^2$. The stability of the
aforementioned solution is analyzed
by introducing the following linear stability ansatz
into Eq. (\ref{gp}):
\begin{equation}
u(x,t)=(u_{0}+\epsilon b) \exp[i((k x-\omega t)+\epsilon a)],
\label{stab}
\end{equation}
where $b$ and $a$ are perturbations of the amplitude and the phase respectively.
By solving the resulting linear equations [to O$(\epsilon)$]
through the Fourier mode expansion $a=a_0 \exp[i (Q x-\eta t)]$ and
$b=b_0 \exp[i (Q x- \eta t)]$, where $Q$ and $\eta$ are the wavenumber and frequency
of the perturbations respectively ($a_0, b_0=$ const.), we finally obtain the dispersion relation 
%
\begin{equation}
(\eta-2 k Q)^2=Q^2 (Q^2+2 g u_0^2).
\label{stab2}
\end{equation}

Noting that the plane wave solution is modullationally stable if and only if
Im$(\eta)=0$,
we will limit our considerations to the focusing case with $g=-1$, which
may potentially lead to MI (apparently, in that case $\eta$ becomes complex).
In this case, Eq. (\ref{stab}) shows that if
$Q<Q_{\rm cr}=u_0 \sqrt{2}$
the plane wave solution becomes modulationally unstable.
So, we notice that there is an implicit dependence on
the size of the domain, $L$. This is, in particular, because the
wavenumbers accessible in a domain of size $L$ are bounded from below
by $Q_{\rm min}=2 \pi/L$. Therefore, we infer (as was also explained in the
introduction) that there exists a minimum $L$, which we denote
\begin{equation}
L_{\rm cr}=\frac{2\pi}{Q_{\rm cr}}=\frac{\sqrt{2}\pi}{u_0},
\label{crit}
\end{equation}
such that $L<L_{\rm cr} \Rightarrow Q_{\rm min}>Q_{\rm cr}$,
and therefore the uniform solution is {\it not} subject
to MI and is thus guaranteed to be dynamically stable.

\subsection{Numerical Experiment I: Expanding Domain, No Trap}

\begin{figure}
\begin{center}
\includegraphics[width=80mm]{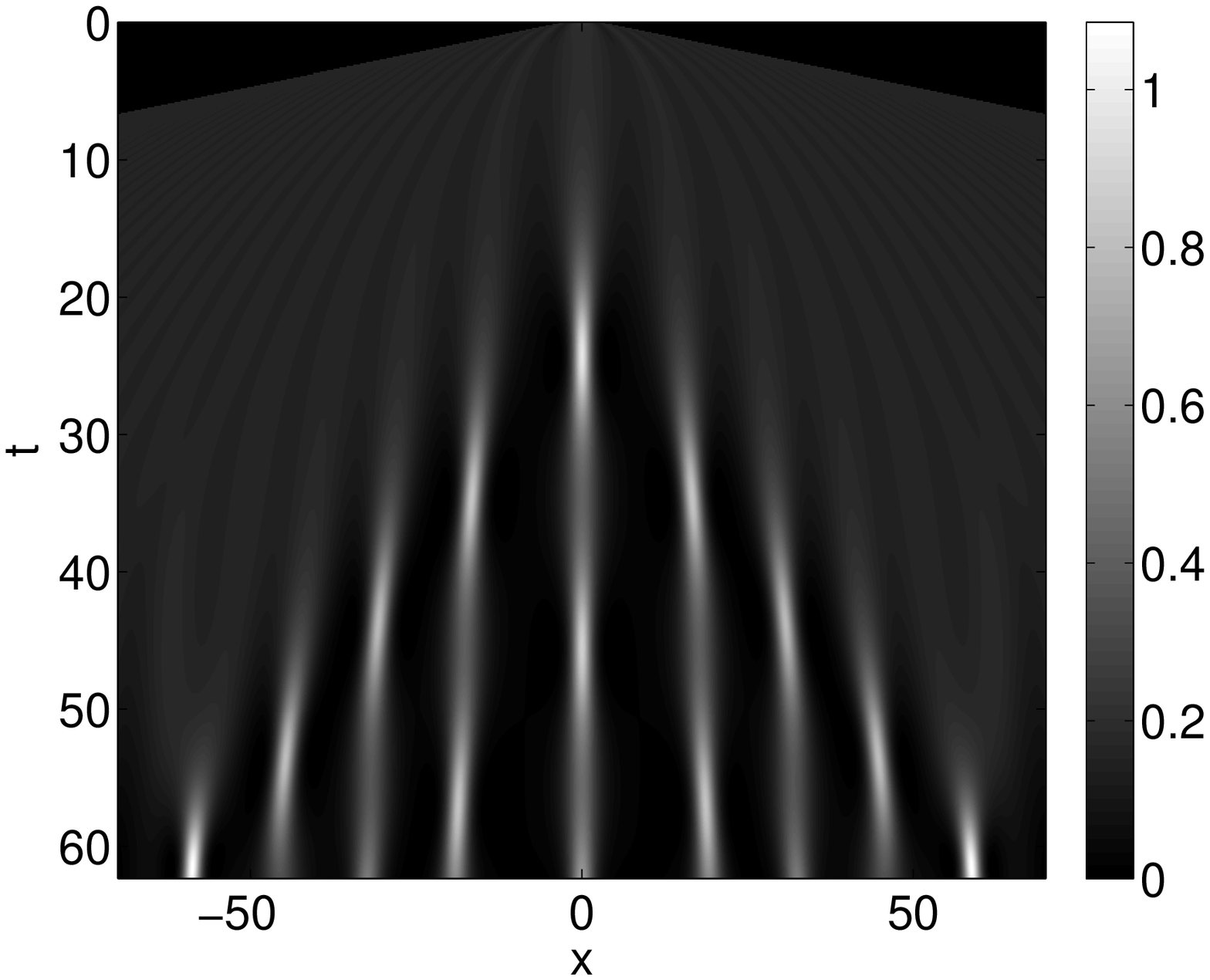}\label{full}
\end{center}
\begin{center}
\begin{tabular}{cc}
\includegraphics[width=80mm]{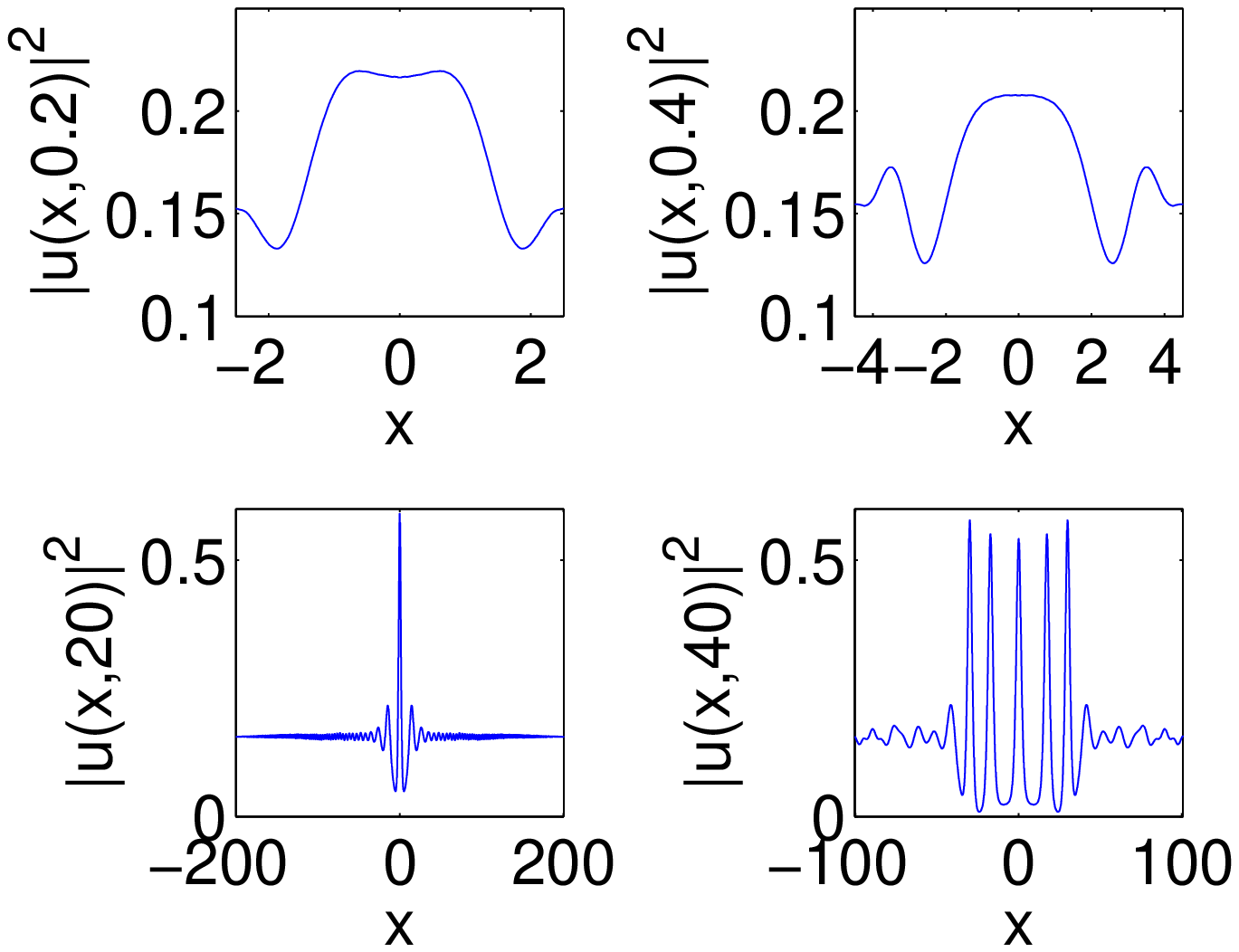}\label{space}
\includegraphics[width=83mm]{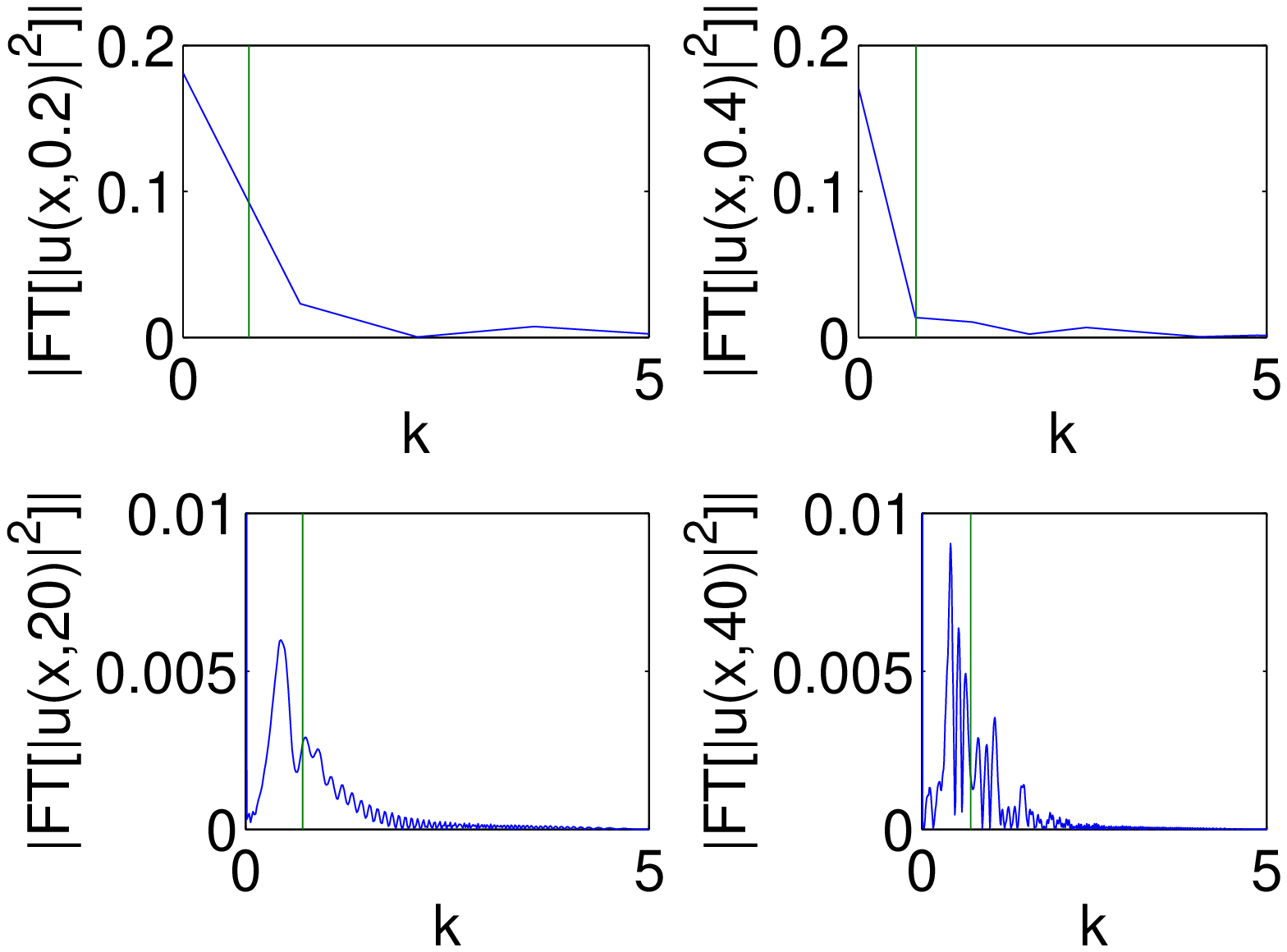}\label{fourier}\\
\end{tabular}
\end{center}
\caption{Top panel: Space-time evolution plot of the field intensity
(square modulus) $|u|^2$ for an initial condition with $u_0=0.5$ and a
perturbation of $0.05 \cos(2{\pi} x)$. Bottom left panels: Four
spatial profiles at times $t=0.2$ (stable), $t=0.4$ (marginal), $t=20$ (unstable) and
$t=40$ (unstable). Bottom right panels: Four respective Fourier profiles;
the smallest wavenumber
present in the solution crosses the critical threshold exactly when the
domain size equals the critical value, $L_{\rm cr}=2\sqrt{2}\pi$. With time, more
subcritical wavenumbers appear and are accordingly amplified.}
\label{kfig1}
\end{figure}

It can be anticipated, based on the above discussion, that upon growing
such a stable domain beyond the critical size $L_{\rm cr}$, unstable wavenumbers may emerge.
So, starting with $L<L_{\rm cr}$, we perturb the plane wave solution of amplitude
$u_0=0.5$ with a stable wavenumber ($Q=2\pi$) and numerically
implement the solution of Eq. (\ref{gp}) in a growing domain $(-l(t),l(t))$,
such that $l(t)=0.5+10t$ (here $L=2l(0)=1$). The resulting evolution
is illustrated in Fig. \ref{kfig1}. The top panel shows the space-time
evolution of the intensity (the solution's square modulus $|u|^2$), illustrating
the emergence of a ``soliton sprinkler'', with an increasing number of
solitary waves appearing, as the domain size increases.
The bottom left four panels show snapshots of the time evolution of
the intensity, clearly indicating the growth of the perturbation and the
eventual formation of multiple solitary wave peaks. Even more
importantly, the bottom right set of panels elucidates the evolution
in Fourier space, with the principal (nonzero) wavenumber initially
being {\it above} the critical wavenumber for the instability (shown
by the vertical line). As time evolves however, $L$ increases, hence
$Q_{\rm min}$ decreases, and upon the crossing of the critical point
we observe growth of the relevant perturbation wavenumbers and
hence the concomitant generation of the localized peaks in real space.


We note that in this numerical experiment, as well as in those that 
will be presented in the following 
subsections,  
we have used a uniform space mesh of size 
$\Delta x=10^{-2}$ and a time step of $h=10^{-5}$.  The methods
employed are centered, second-order finite difference for 
spatial discretization and fourth-order Runge-Kutta for time integration.
We use periodic boundary conditions in order to effectively capture the frequency decomposition.

In the present implementation, rather than rescaling space as 
was done in some of the earlier works in the dissipative context,
we maintain the same mesh-size, while {\it actually}
increasing the domain with extra nodes at the endpoints of the domain.
Notice that the above figure (Fig. \ref{kfig1}) doesn't depict the entire final
domain, which is approximately $(-600,600)$, but merely the relevant activity in the center.



\subsection{Numerical Experiment II: Contracting Domain, No Trap}

We now turn to a less orthodox
manifestation of MI, which, however, illustrates an important additional
dependence of the instability's critical point, namely the dependence
on the amplitude. If, instead of increasing the domain size, we
decrease the domain size while increasing background intensity
level, so as to maintain the conservation of the $L^2$ norm, we
also observe a manifestation of MI.

\begin{figure}
\begin{center}
\includegraphics[width=80mm]{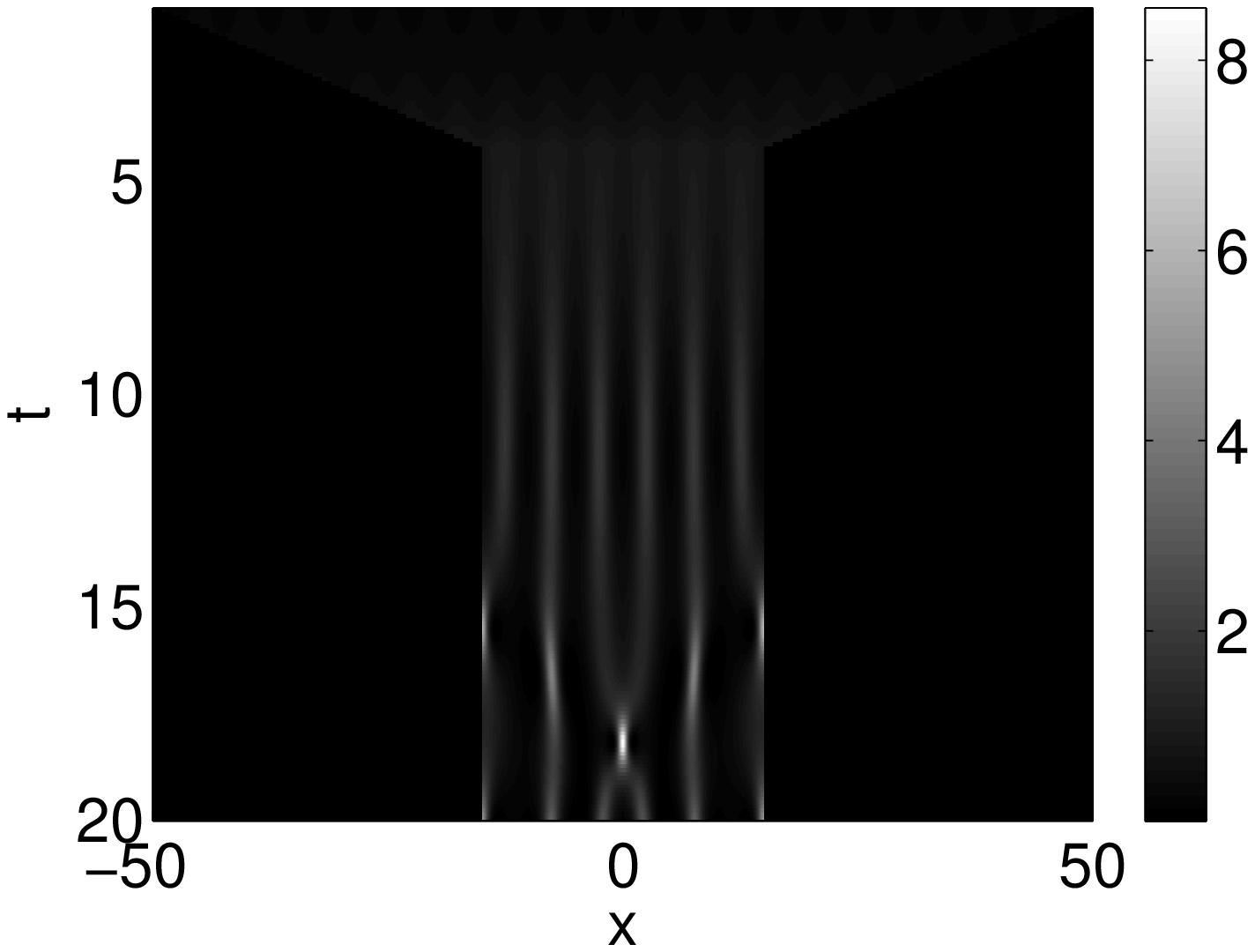}
\end{center}
\includegraphics[width=80mm]{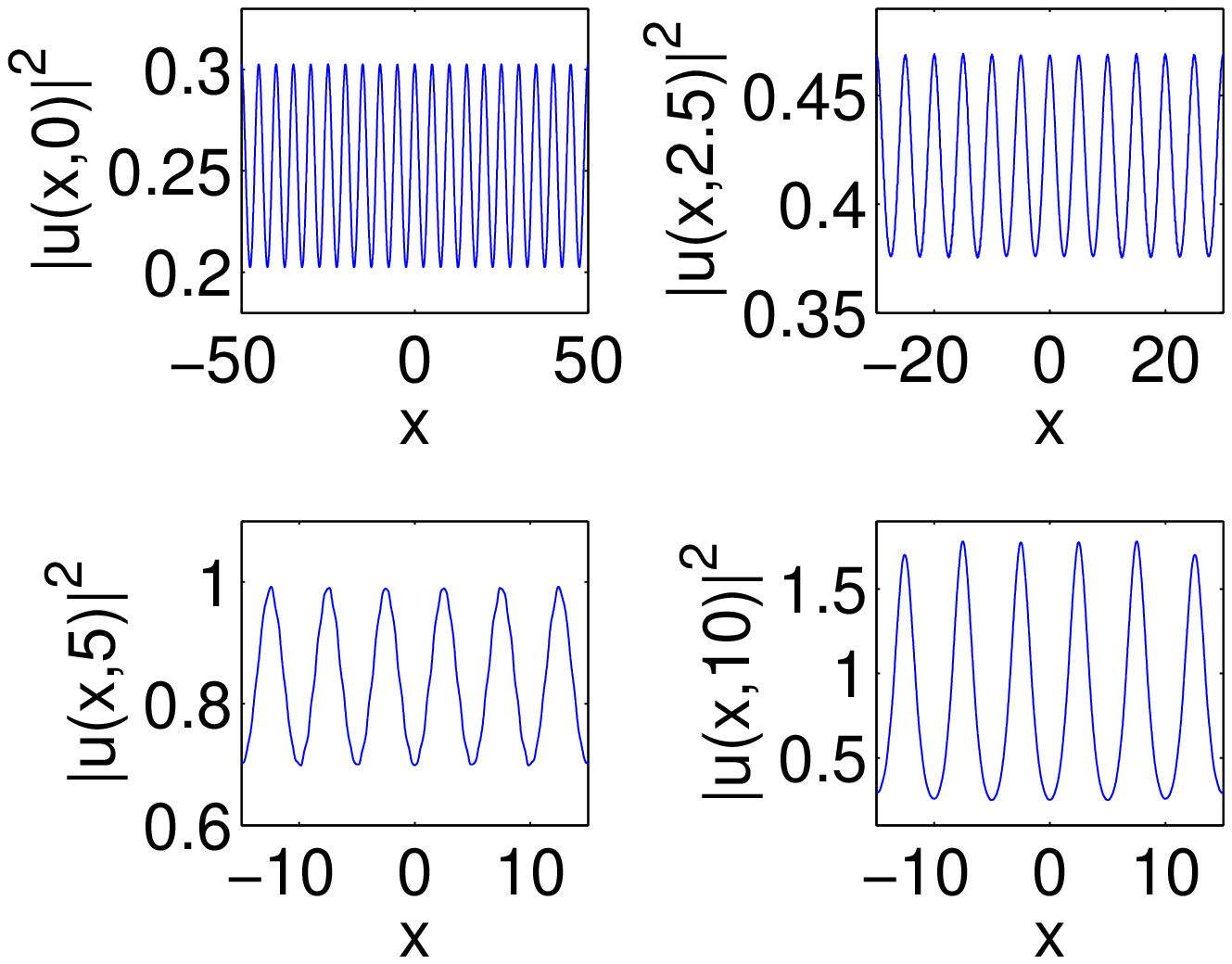}
\includegraphics[width=80mm]{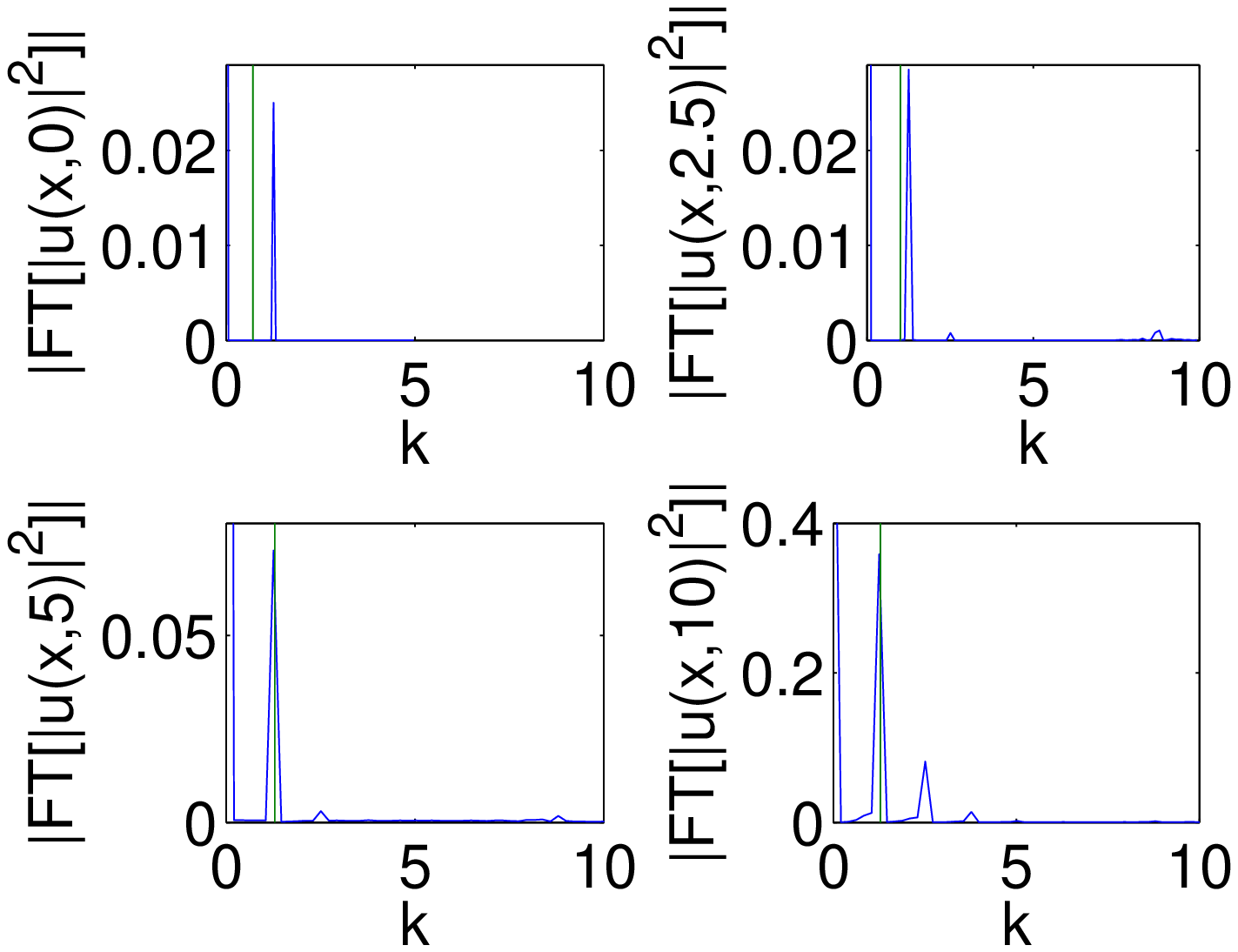}
\label{decrease}
\caption{Same as Figure \ref{kfig1}, but for the decreasing domain case
with fixed $L^2$ norm.
(Top) The emergence of instability in a space-time intensity plot
with decreasing domain size beginning with $l(0)=50$, $K=10$, ${\psi_0}=0.5$,
and a perturbation of $0.05\cos(2{\pi} x/5)$. Bottom left: Four typical
spatial profiles
before (upper) and after (bottom) the manifestation of the instability.
Bottom right: The four corresponding Fourier profiles are shown, illustrating
the emergence of the instability.}
\label{kfig2}
\end{figure}

In particular, let us consider a plane wave of amplitude $u_1$ in the finite
interval $L^{(1)}$. Then, decreasing the interval to $L^{(2)}<L^{(1)}$,
while conserving the squared 
$L^2$ norm $\int |u|^2 dx$, we obtain a plane wave of amplitude $u_2$,
which is determined by the equation
$|u_1|^2 L^{(1)} = |u_2|^2 L^{(2)}$.
To this end,
utilizing Eq. (\ref{crit}), we readily obtain
%
\begin{equation}
L_{\rm cr}^{(2)}=
L_{\rm cr}^{(1)}\sqrt{\frac{L^{(2)}}{L^{(1)}}} < L_{\rm cr}^{(1)}.
\label{crit2}
\end{equation}
%
The above result shows that $Q_{\rm cr}^{(2)}>Q_{\rm cr}^{(1)}$, which implies that such a
shrinking of the domain (while maintaining conservation of the $L^2$ norm) has
the potential to result in instability of previously stable modes.
In fact, for any $\tilde{L}$, we have a corresponding
$\tilde{Q}_{\rm cr}=\tilde{u} \sqrt{2}= u_1 \sqrt{2L^{(1)}/\tilde{L}}$. 
So, as the
domain size decreases, the critical wavenumber increases, and eventually we
expect it to surpass a wavenumber excited in the solution, resulting in MI.
This is clearly shown in Fig. \ref{kfig2}, which is similar to Fig. \ref{kfig1}, but now for a shrinking domain.
The dynamical procedure involves linearly decreasing the domain $(-l(t),l(t))$
with a rate $dl/dt =-K$ and preserving the $L^2$ norm of the modulus numerically via the
rescaling transformation
\begin{equation}
\int_{-l(t)}^{l(t)} |u|^2 \,dx = \int_{-l_0}^{l_0} |u|^2 \,dx.
\end{equation}






We observe that in Fourier space it is now the critical
point that is shifting to the right, eventually making the configuration
unstable, through its crossing of the perturbation wavenumber. The
resulting amplification induced by the instability is observed both
in the real space formation of large amplitude structures and in the
Fourier space growth of the relevant peaks. A noteworthy feature of
the top panel of Fig. \ref{kfig2} is that at $t \approx 18$ the middle two
localized structures 
undergo a nearly elastic collision which underscores their solitary wave 
character.

\subsection{Numerical Experiment III: Expanding Domain, Parabolic Trap}

We now turn our attention to the more general inhomogeneous case and consider the
full GP equation (\ref{gp}) incorporating the external potential.
The latter
is known to be particularly relevant to the confinement of the
condensate, and assumes typically the harmonic form
$V_{\rm ext}(x)=(1/2)\Omega^2 x^2$, as, e.g., in the case of a magnetic or an optical harmonic trap
of normalized strength $\Omega$ \cite{BEC}.

Our motivation for considering this example is the following.
For Eq. (\ref{gp}) with the focusing nonlinearity and the parabolic potential,
we can obtain the ground state of the system starting, e.g.,
from the simple linear limit of $g=0$. In that case, Eq. (\ref{gp}) is actually
the usual linear Schr\"{o}dinger equation, which possesses
the exact solutions $u(x,t)=\exp(-i \mu t){\psi}(x)$,
where $\psi(x)=\exp(-\Omega x^{2}/ 2 \sqrt{2}) H_{n}(x)$, where $H_{n}(x)$ is
the $n$-th degree Hermite
polynomial and $\mu=(2n+1)\Omega/\sqrt{2}$ is the normalized chemical potential.
One can then straightforwardly
establish (based on continuation arguments), that there is a nonlinear
continuation of this linear state bifurcating from the limit of vanishing
$L^2$ norm (i.e., vanishing number of atoms). This branch of solutions
can be obtained with a fixed point algorithm such as a
Newton method. There is a unique such continuation for every $n$;
see, e.g., \cite{kivshar}. Here we will consider the ground state,
emanating from the solution with $n=0$. For large amplitude, this solution
approaches the well-known soliton solution. This is natural
since here the nonlinearity tends to focus the solution towards
the center of the 
harmonic trap, where $V_{\rm ext}(x) \rightarrow 0$ and, hence,
we revert to the solitonic solution of the untrapped system.

However, for small amplitudes, this ground state branch represents a
more extended, nearly-linear solution (i.e., a weakly nonlinear generalization
of the linear ground state in the presence of the trap). As such, its
width is still critically determined by the trap strength $\Omega$.
Therefore, by changing this parameter
we may impose on such a nearly-linear solution an effective
change in the size of the domain accessible to it. In so doing,
we can produce an effectively expanding domain, by virtue of decreasing
$\Omega$, or an effectively contracting domain, by means of increasing
$\Omega$. These are,
respectively, the types of numerical experiments that we will present
in this and in the following subsections.

\begin{figure}
\begin{center}
\begin{tabular}{ccc}
\includegraphics[width=55mm]{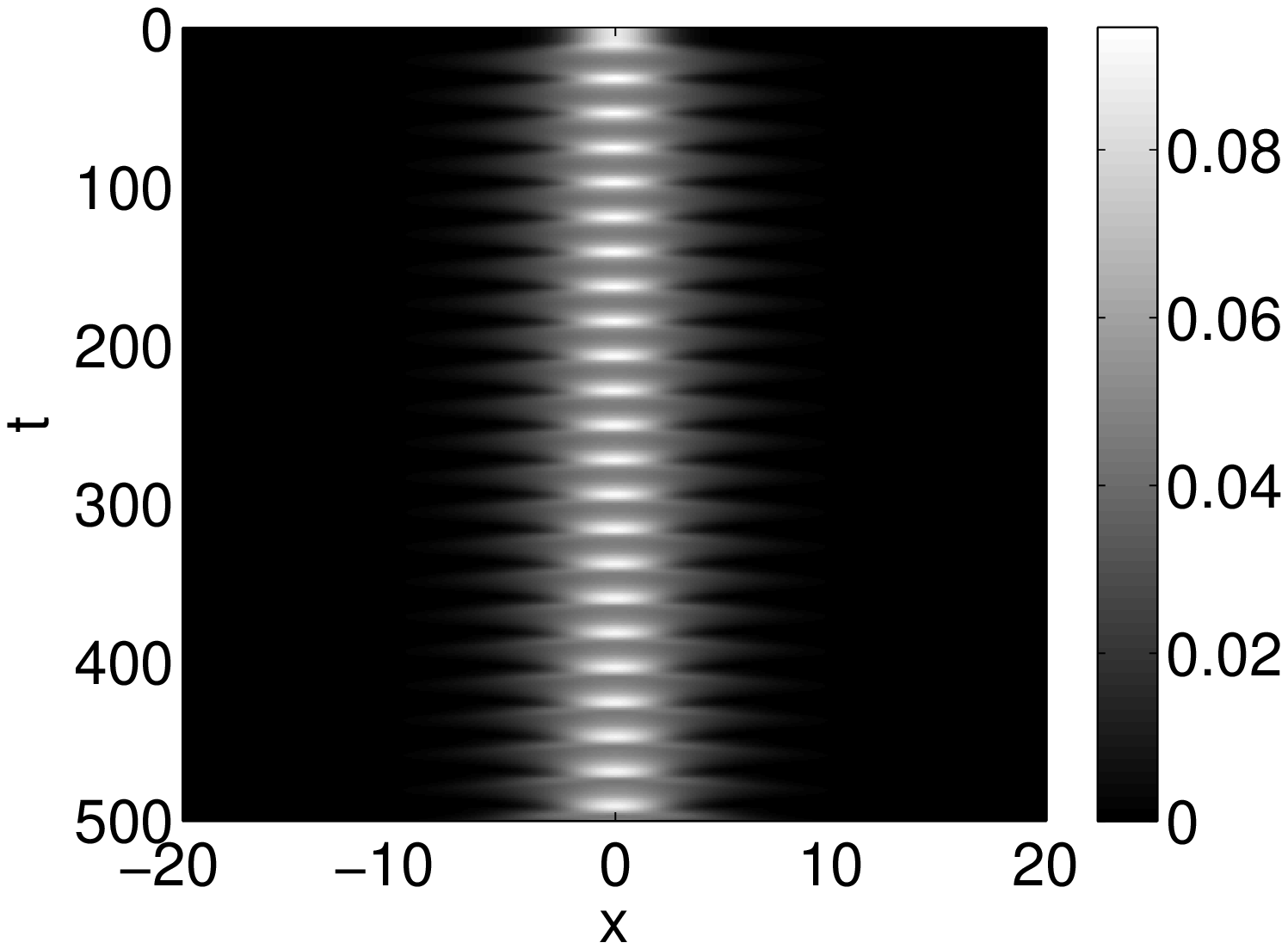} &
\includegraphics[width=55mm]{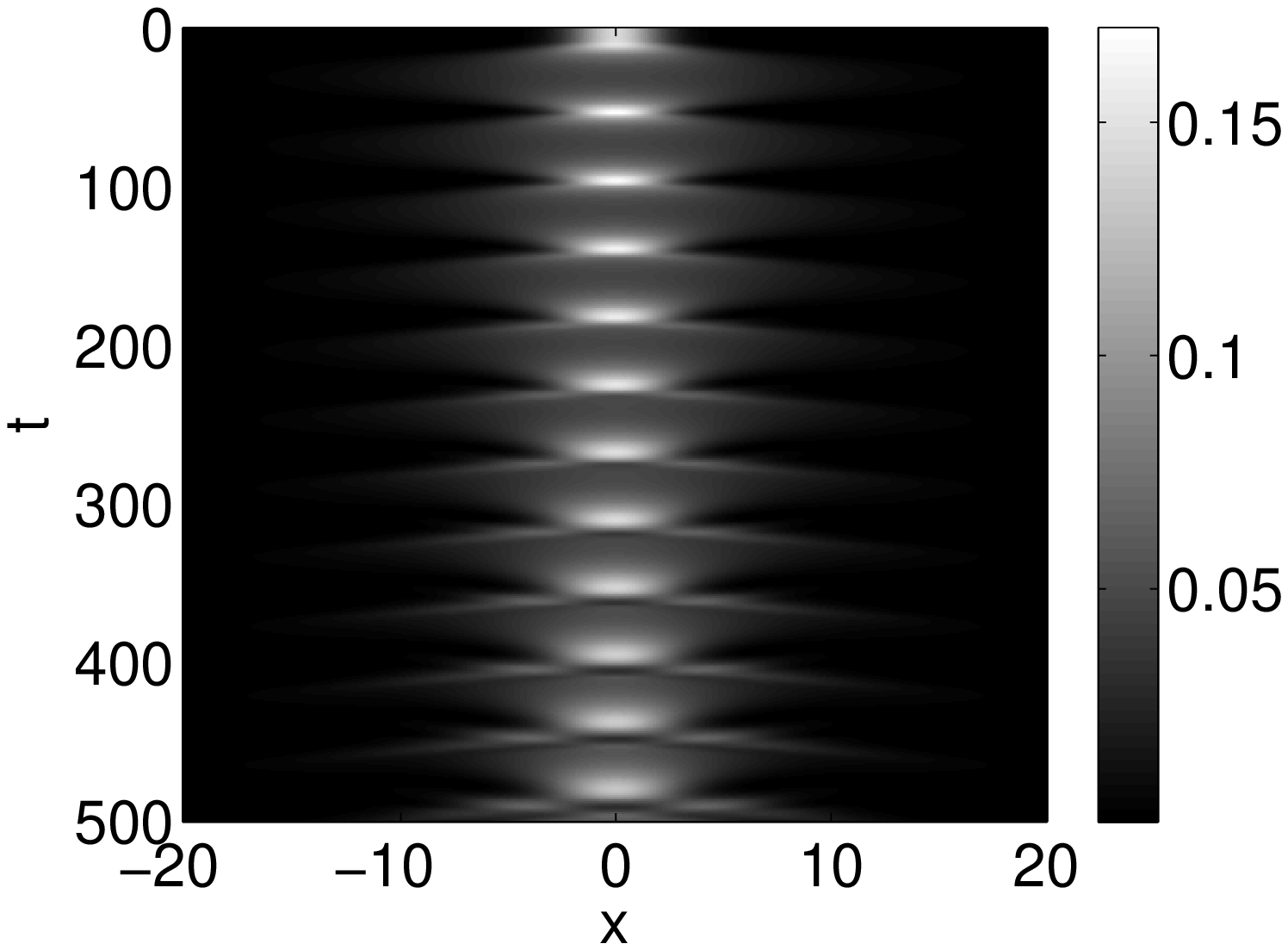} &
\includegraphics[width=55mm]{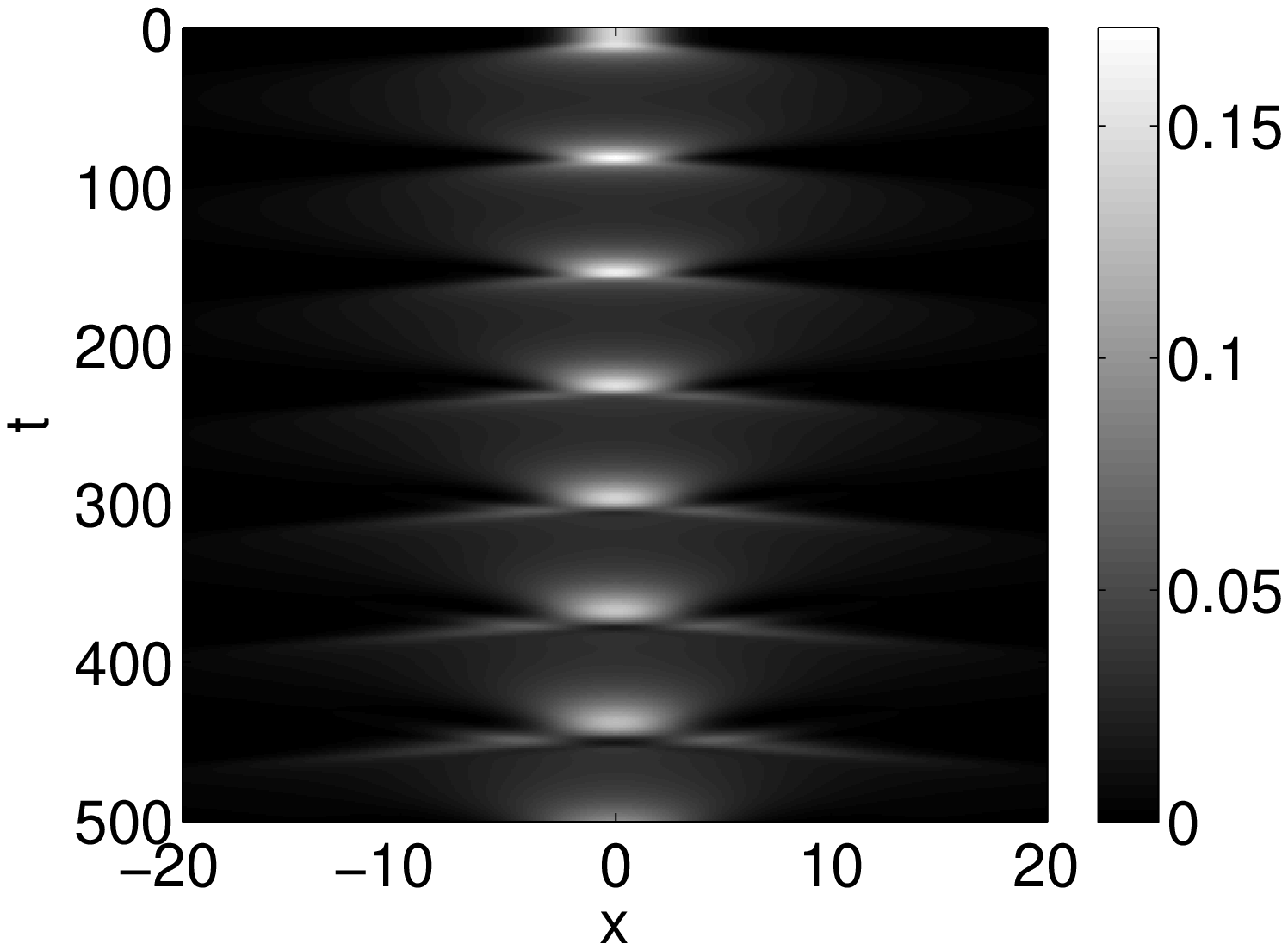} \\
\includegraphics[width=55mm]{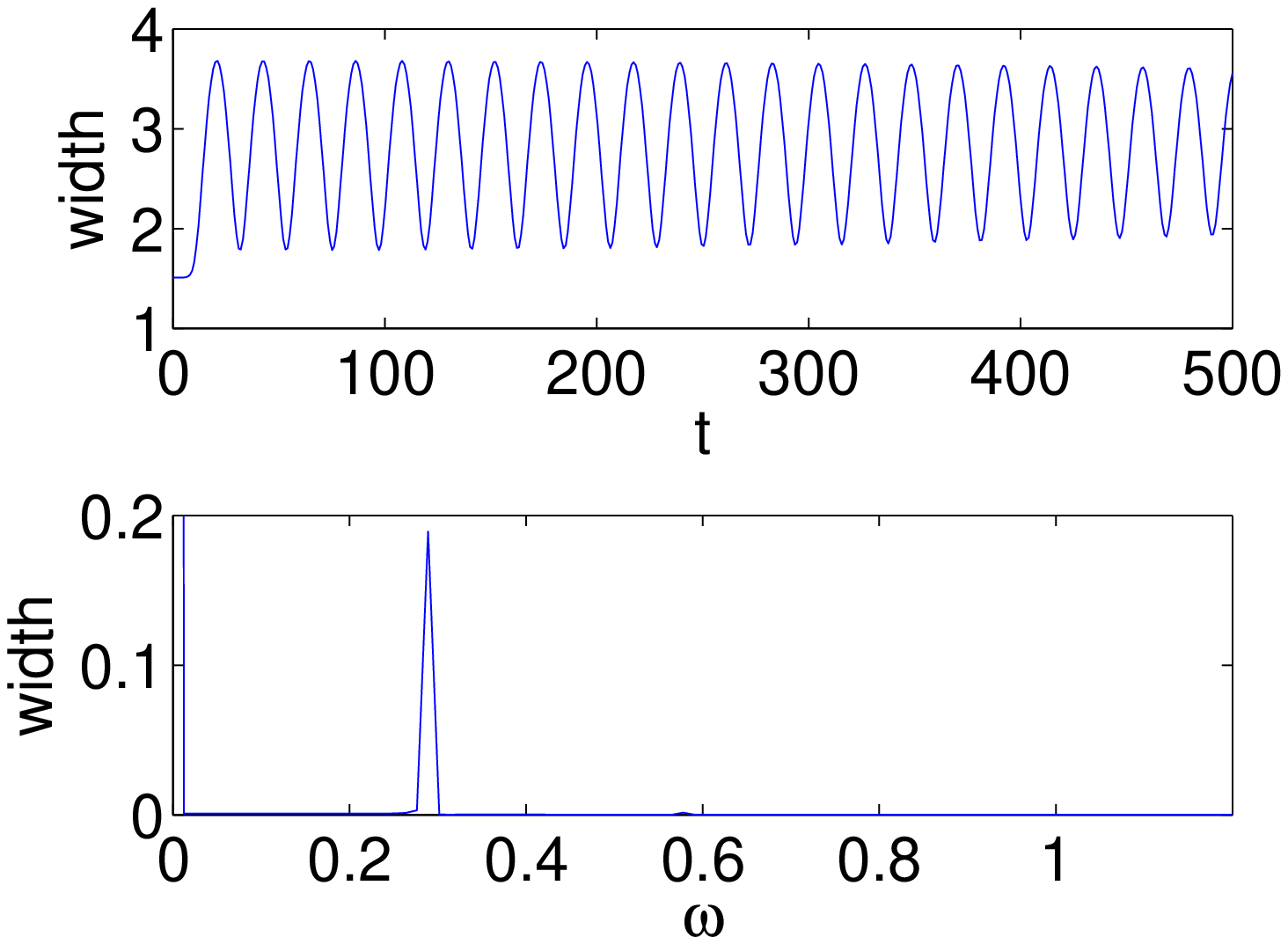} &
\includegraphics[width=55mm]{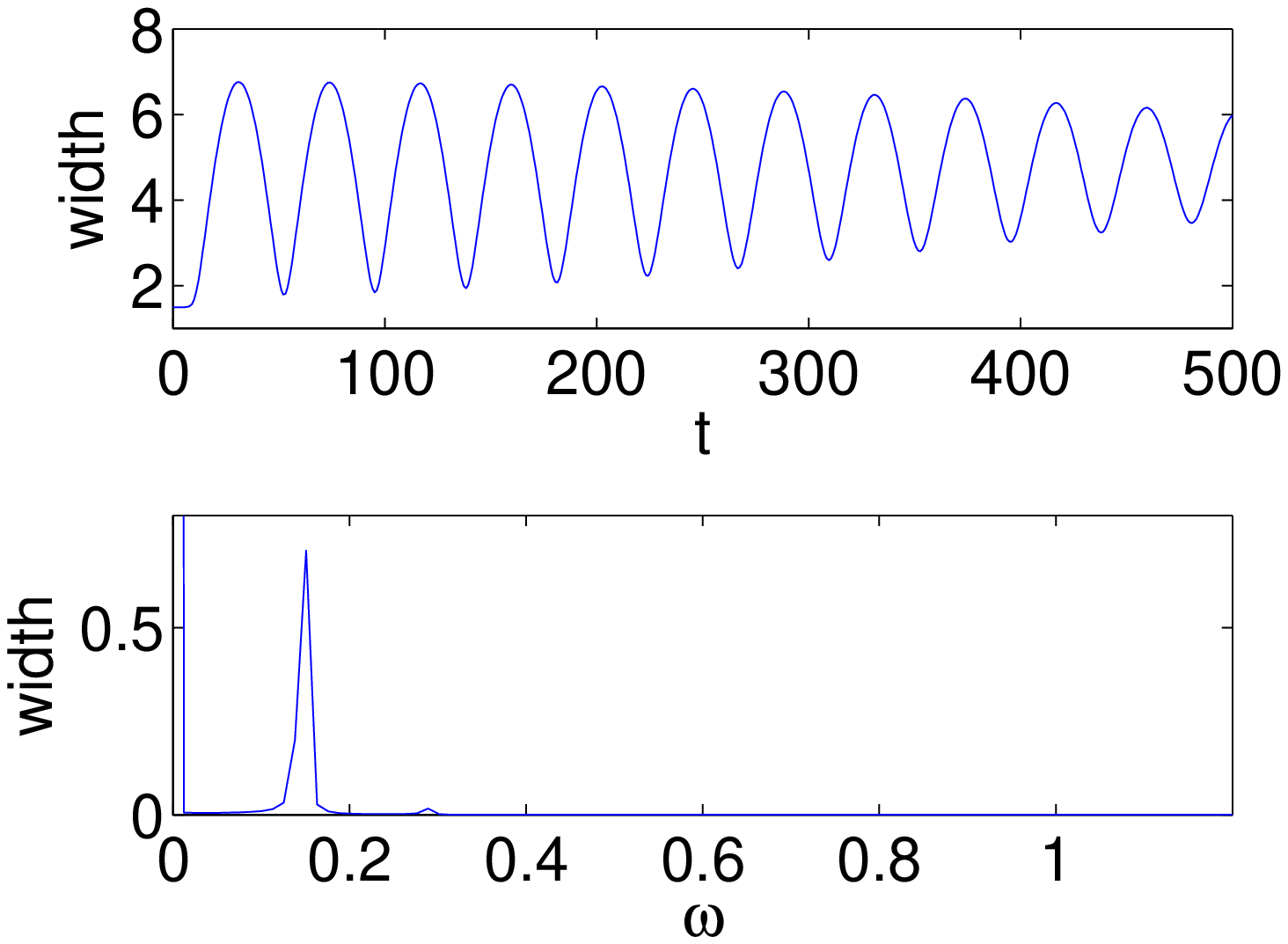} &
\includegraphics[width=55mm]{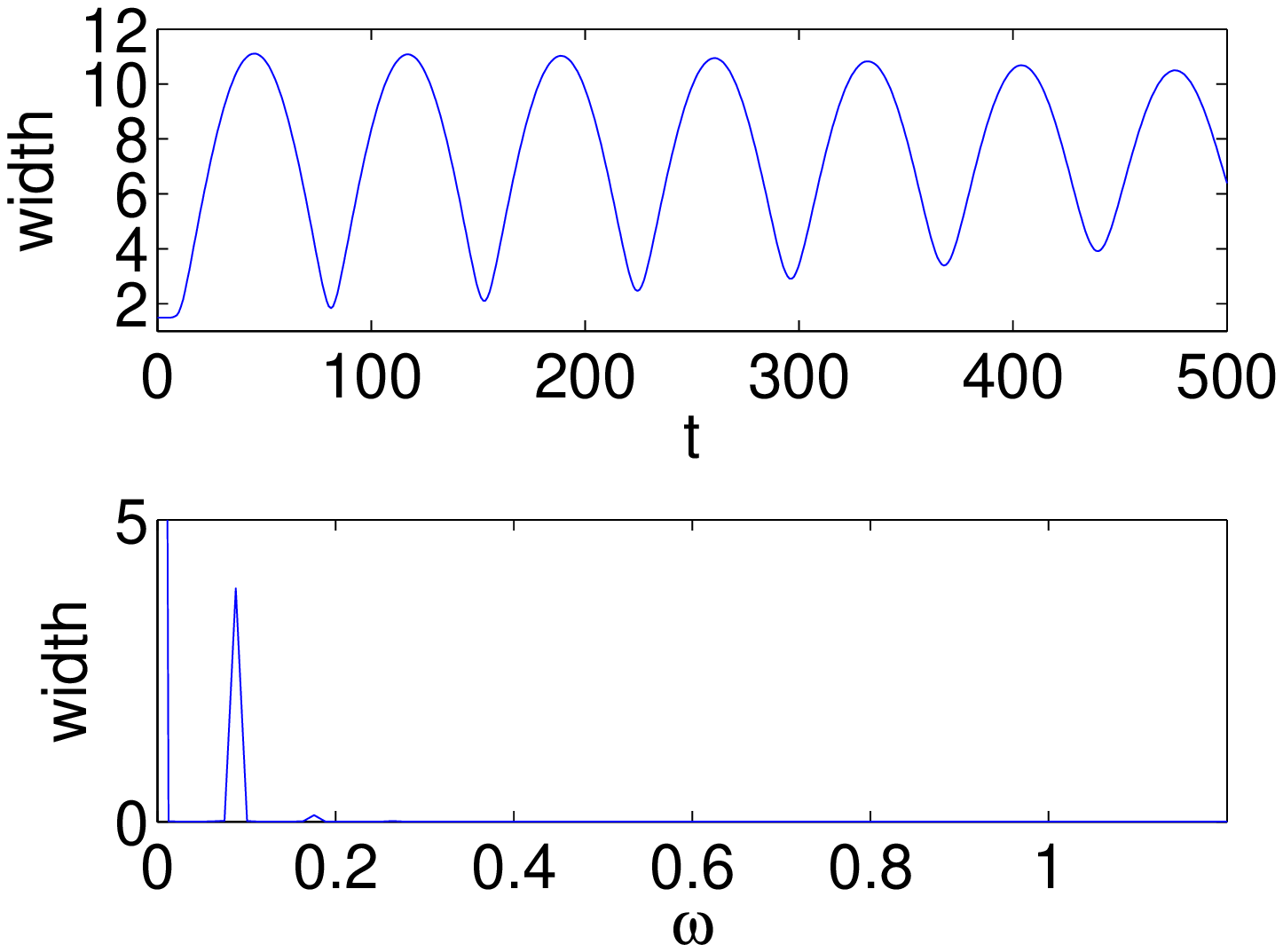} \\
\includegraphics[width=55mm]{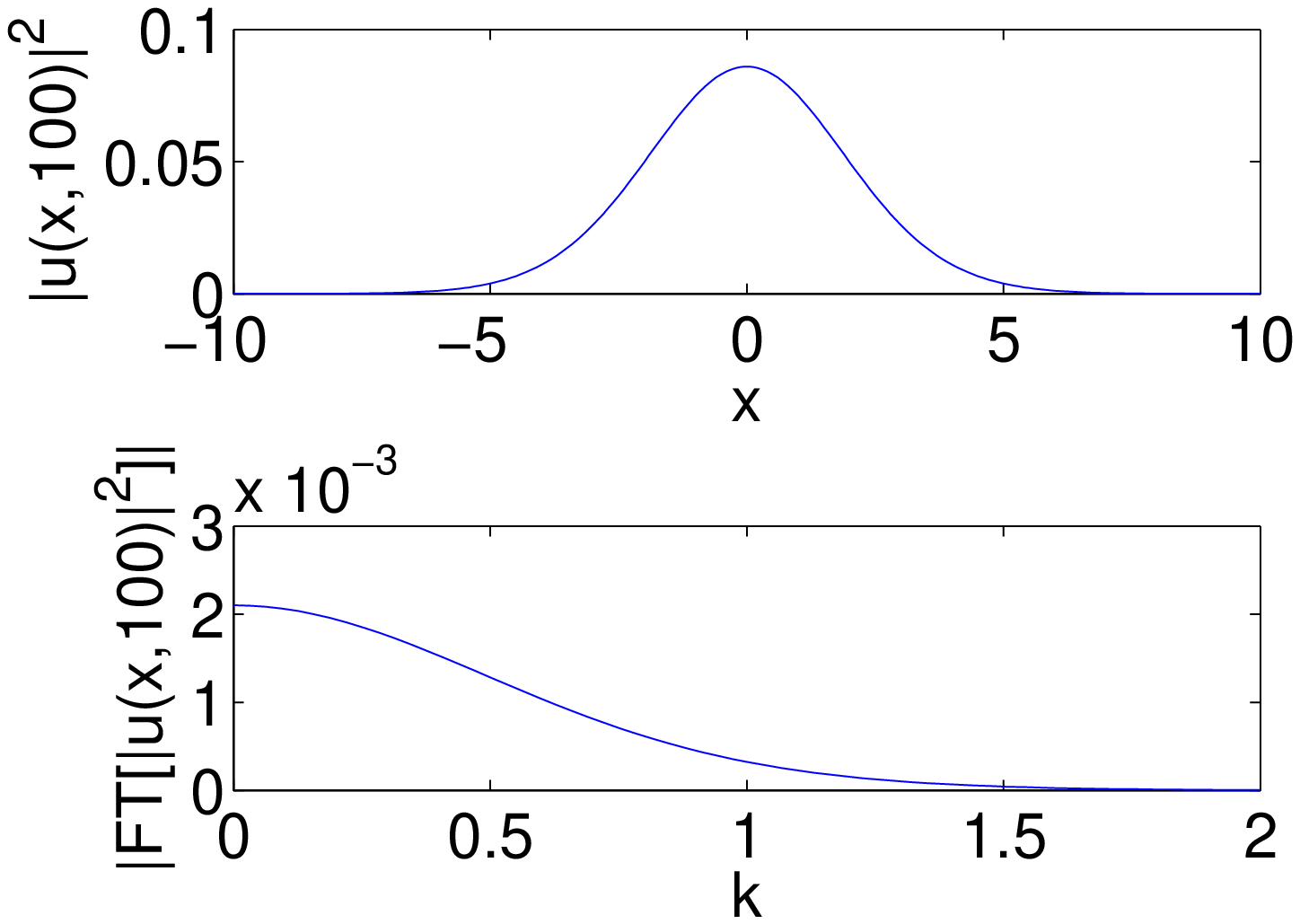} &
\includegraphics[width=55mm]{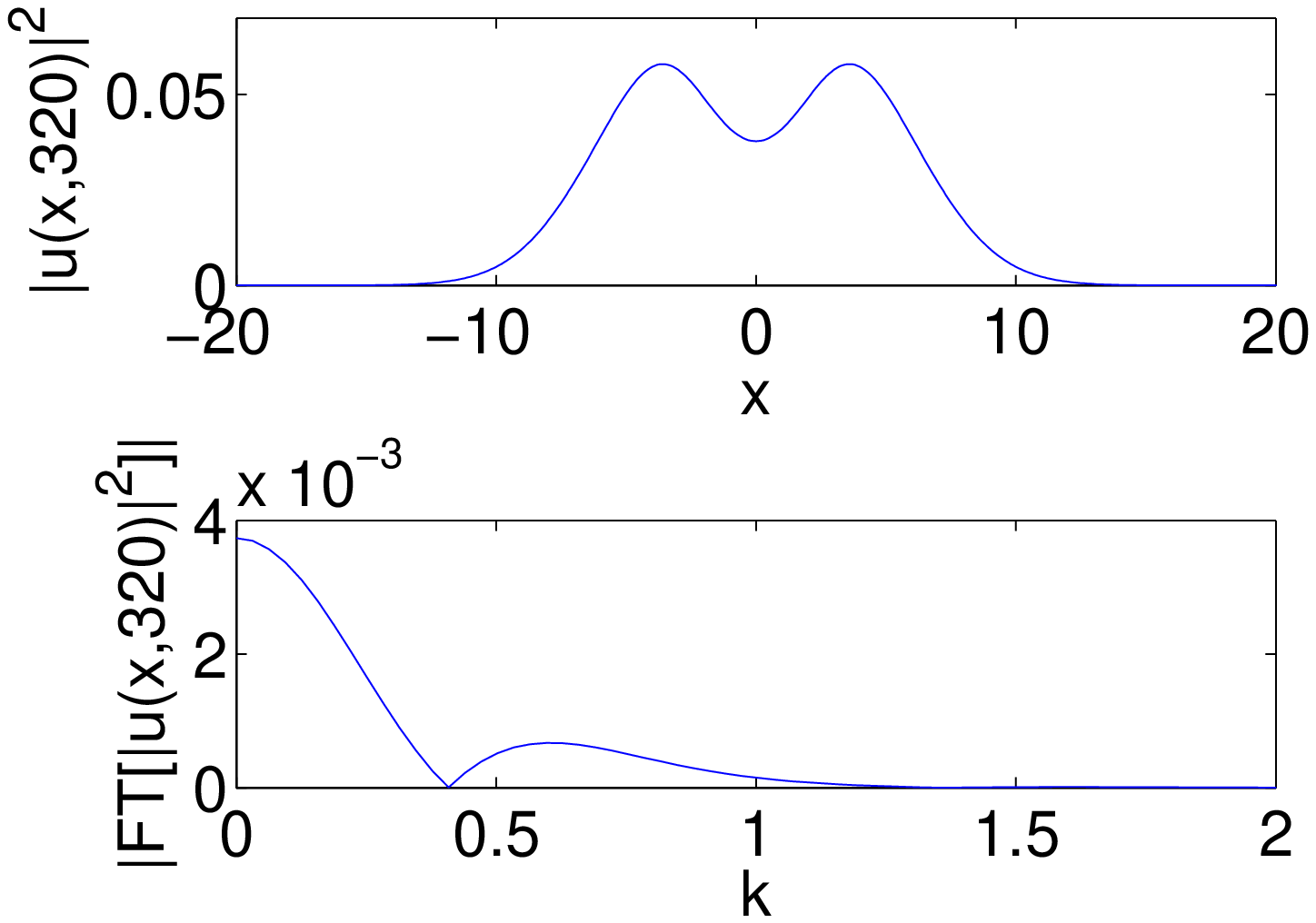} &
\includegraphics[width=55mm]{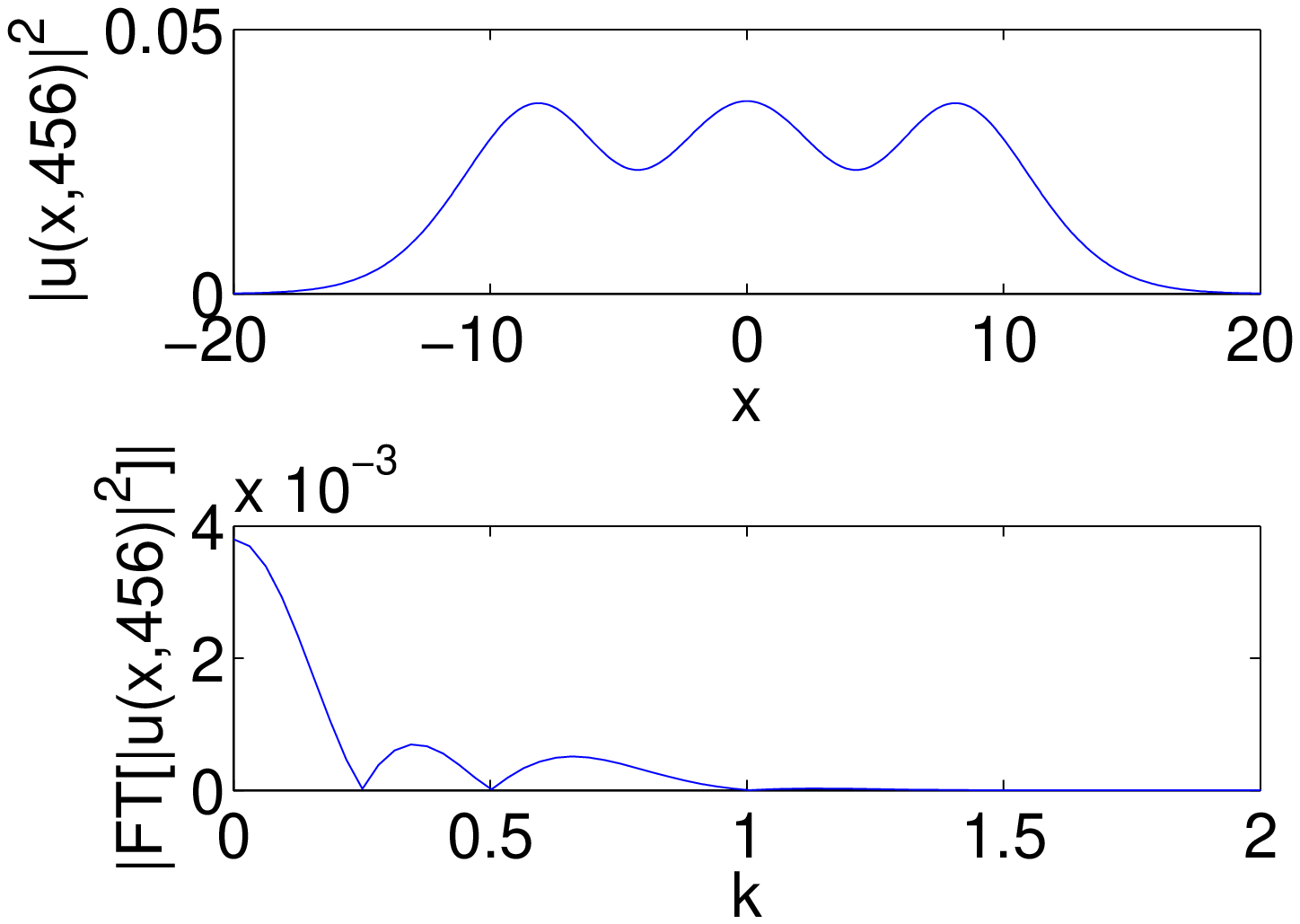} 
\end{tabular}
\end{center}
\caption{Top row: Intensity space-time contour plots with
$\Omega_1=0.3$, $\Omega_2=0.1$ (left), $\Omega_2=0.05$ (middle), and
$\Omega_2=0.03$ (right). Middle two rows: The corresponding widths
as a function of time (upper panels) and frequencies (lower panels).
Bottom two rows: Sample spatial profiles (upper panels) and
corresponding Fourier profiles (lower panels).} \label{kfig3}
\end{figure}

In order to modify the harmonic
trap strength, we use a time-dependent parameter $\Omega$,
according to
\begin{equation}
{\Omega}(t)={\Omega}_1+({\Omega}_2-{\Omega}_1)(1+\tanh[(t-t_0)/{\tau}])/2, 
\label{omega}
\end{equation}
%
i.e., for $t \ll t_0$, the trap strength
is $\Omega \approx \Omega_1$, for $t \gg t_0$, it is $\Omega \approx
\Omega_2$ (and hence we can expand or shrink the domain, depending on
whether $\Omega_2<\Omega_1$ or $\Omega_2>\Omega_1$, respectively).

Starting from the case of expanding the domain, we show the
relevant results in Figure \ref{kfig3}. However, in this version
of the problem (i.e., with the parabolic trap), when simply expanding
the domain the amplitude of the solution will decrease (rather than
being the same as is the case at the boundaries in Experiment I).
However, in order to observe the instability, the amplitude of the solution
should be preserved. This is achieved by an ``injection'' of matter
in the system, so as to preserve the original amplitude.


This is accomplished in a multiple step process in which we
initially integrate until the end of the ramp, at approximately
$t^*=t_0+3\tau$. Next, we calculate the amplitude loss as $a =
{\rm max}_x(|u(0)|)- {\rm max}_x(|u(t^*)|)$.  Finally we rerun the
simulation, except this time we augment the right hand side of Eq. (\ref{gp}) 
with a spatially uniform gain rate $\zeta(t)$ over the time when the trap opens up. 
In particular, the considered modified GP equation and the gain term 
assume, respectively, the forms:
\begin{eqnarray}
i \partial_t u &=& 
-\partial_{x}^{2} u + g|u|^2 u + V_{\rm ext}(x)u +i \zeta(t) u 
\label{mgpe} \\
\zeta(t) &=&  \frac{a}{2 {\tau}}{\rm sech}^2[(t-t_0)/{\tau}], 
\label{gain}
\end{eqnarray}
which may model the employement of a source of BEC atoms, as in the experiment \cite{as}.
We then observe that if the trap strength is decreased slightly (left set of panels
in Fig. \ref{kfig3}), the dynamics
is going to be stable, in the sense that the distribution of matter
will remain unimodal and its width will perform simple
small-amplitude oscillations, while the Fourier space profile will
remain monotonic (quasi-Gaussian, as the Fourier transform
of a near-Gaussian profile in real space). On the other hand, if
the trap strength is decreased significantly (middle and right set of
panels in Figure \ref{kfig3}), then the 
system becomes ``unstable'', in the sense that the spatial profile
distribution will become multi-modal (in fact with more humps developing for smaller frequencies), the oscillations
of the matter wave width will become more complex (and involve more
frequencies) and the Fourier profile will develop nodes, 
and emergent bands of higher wavenumber excitations, whose number
is higher the more multi-modal the spatial distribution becomes.

\subsection{Numerical Experiment IV: Contracting Domain, Parabolic Trap}

 \begin{figure}
\begin{center}
\begin{tabular}{ccc}
\includegraphics[width=55mm]{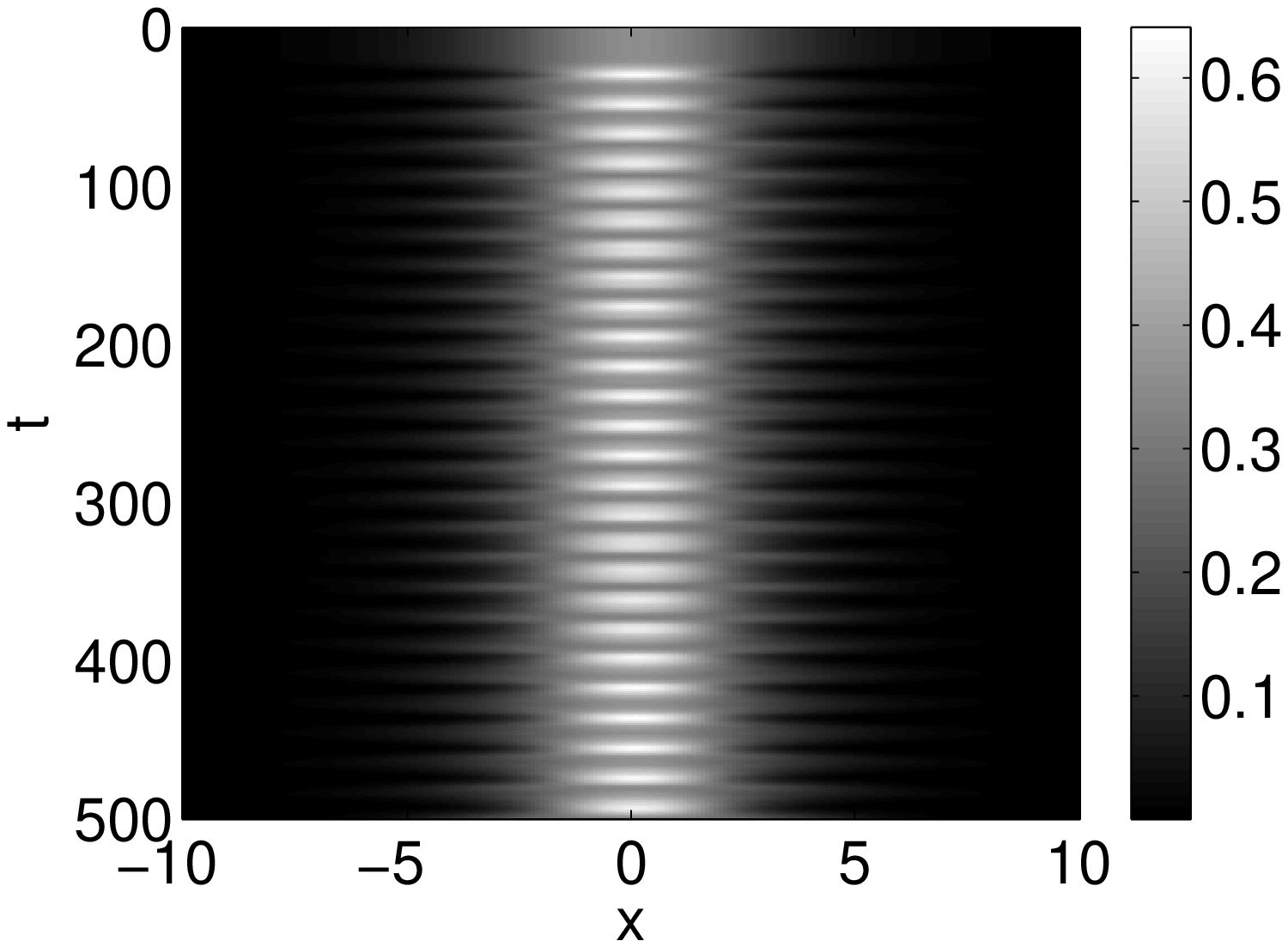} &
\includegraphics[width=55mm]{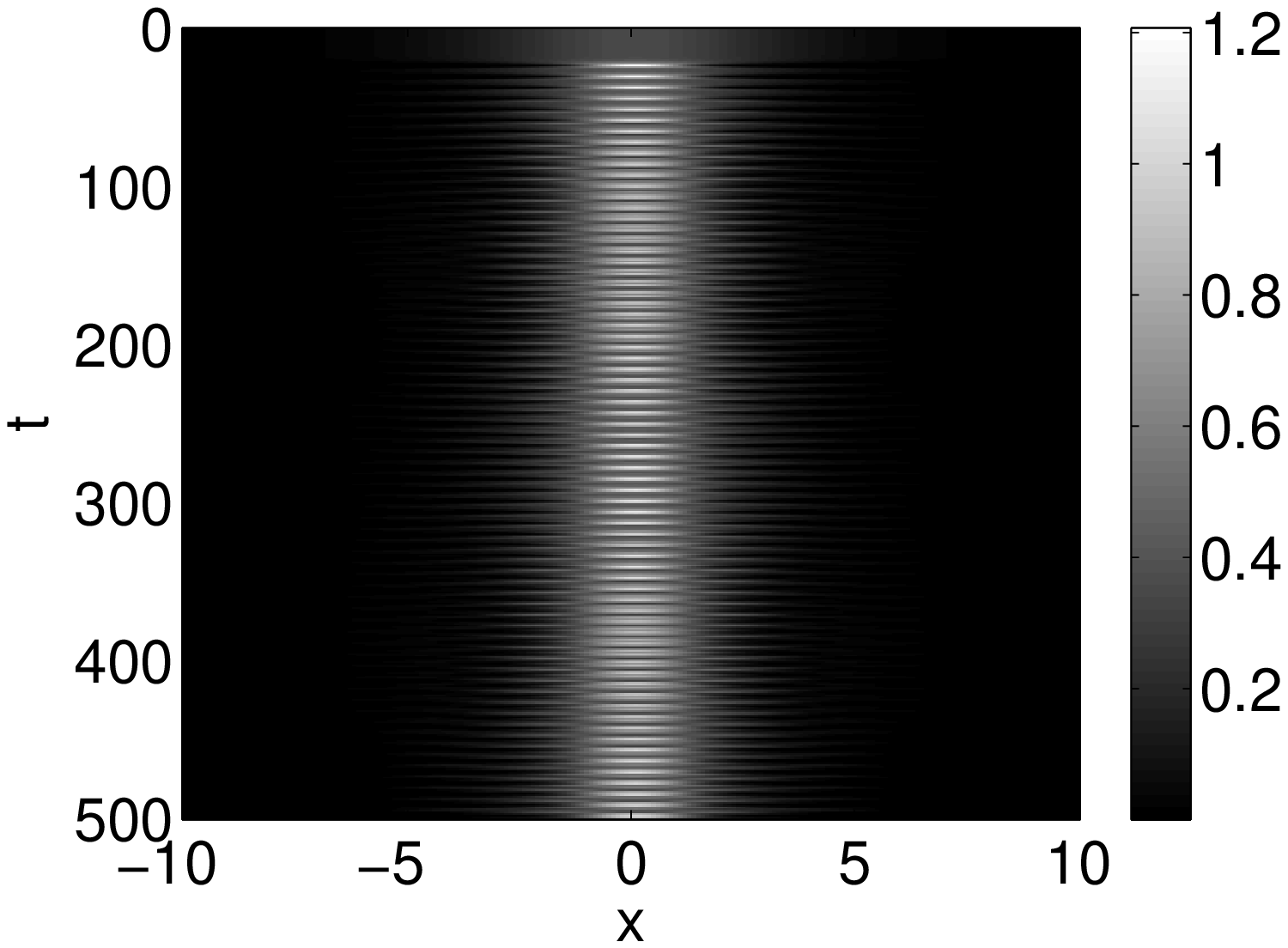} &
\includegraphics[width=55mm]{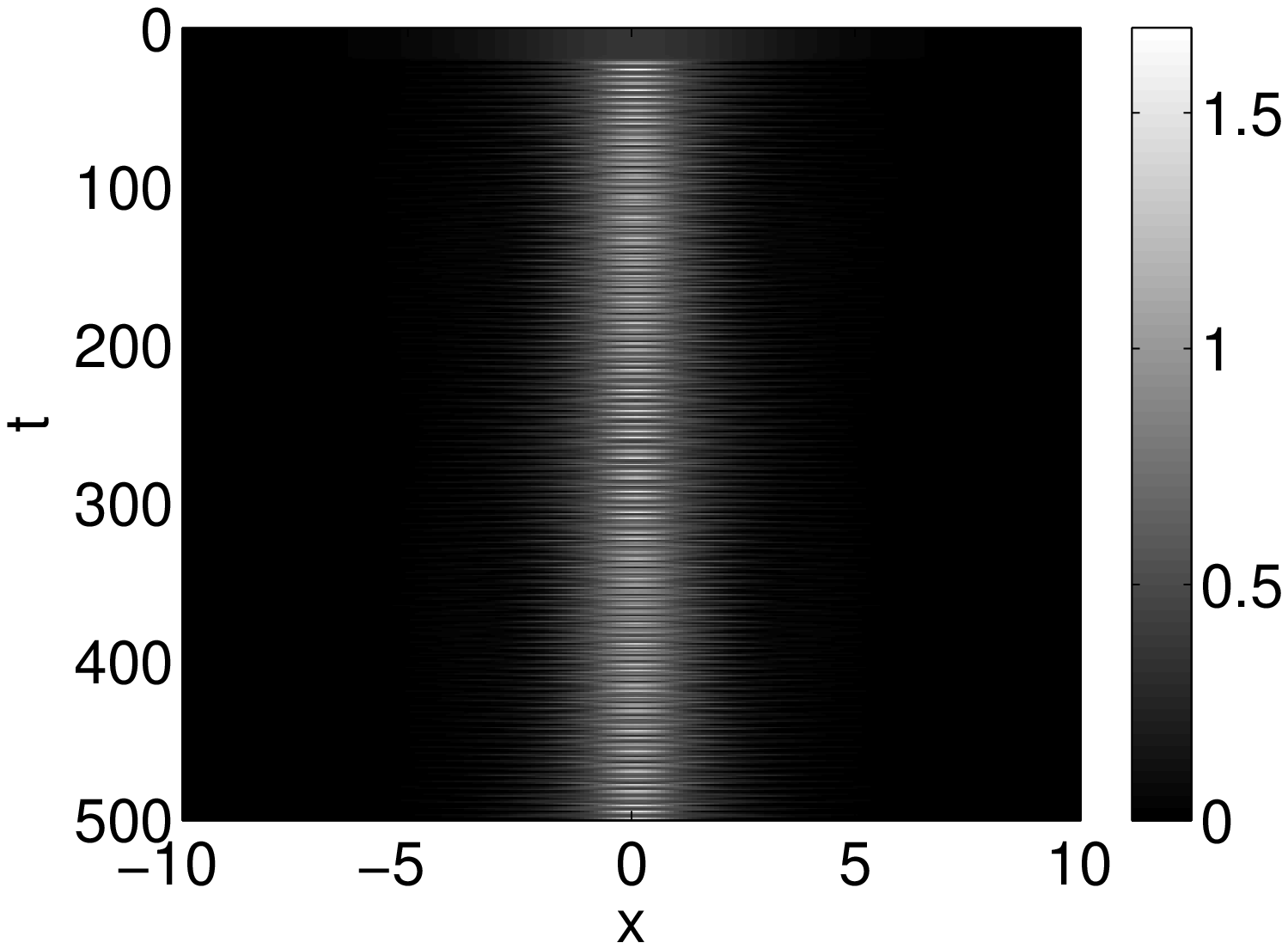} \\
\includegraphics[width=55mm]{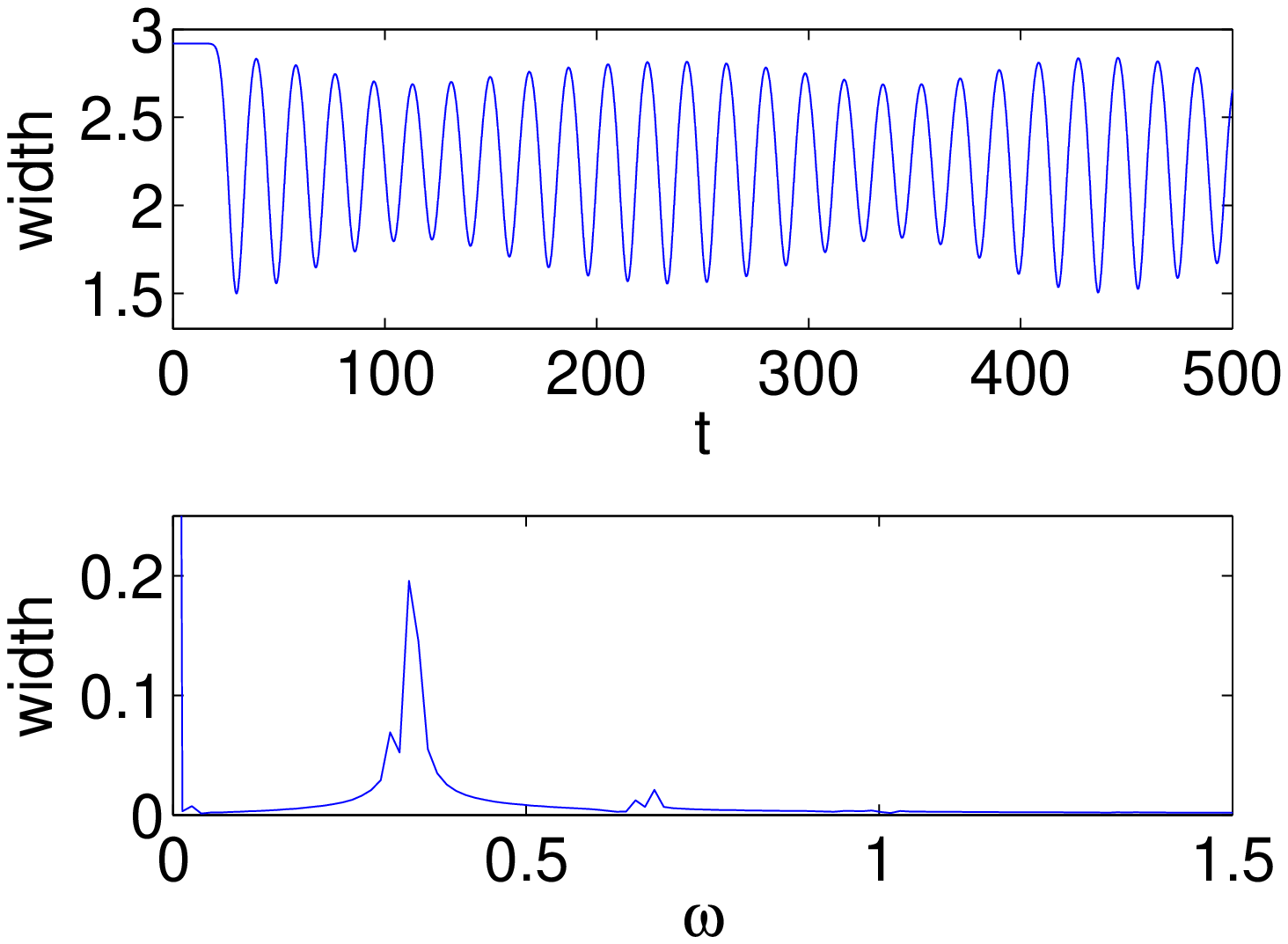} &
\includegraphics[width=55mm]{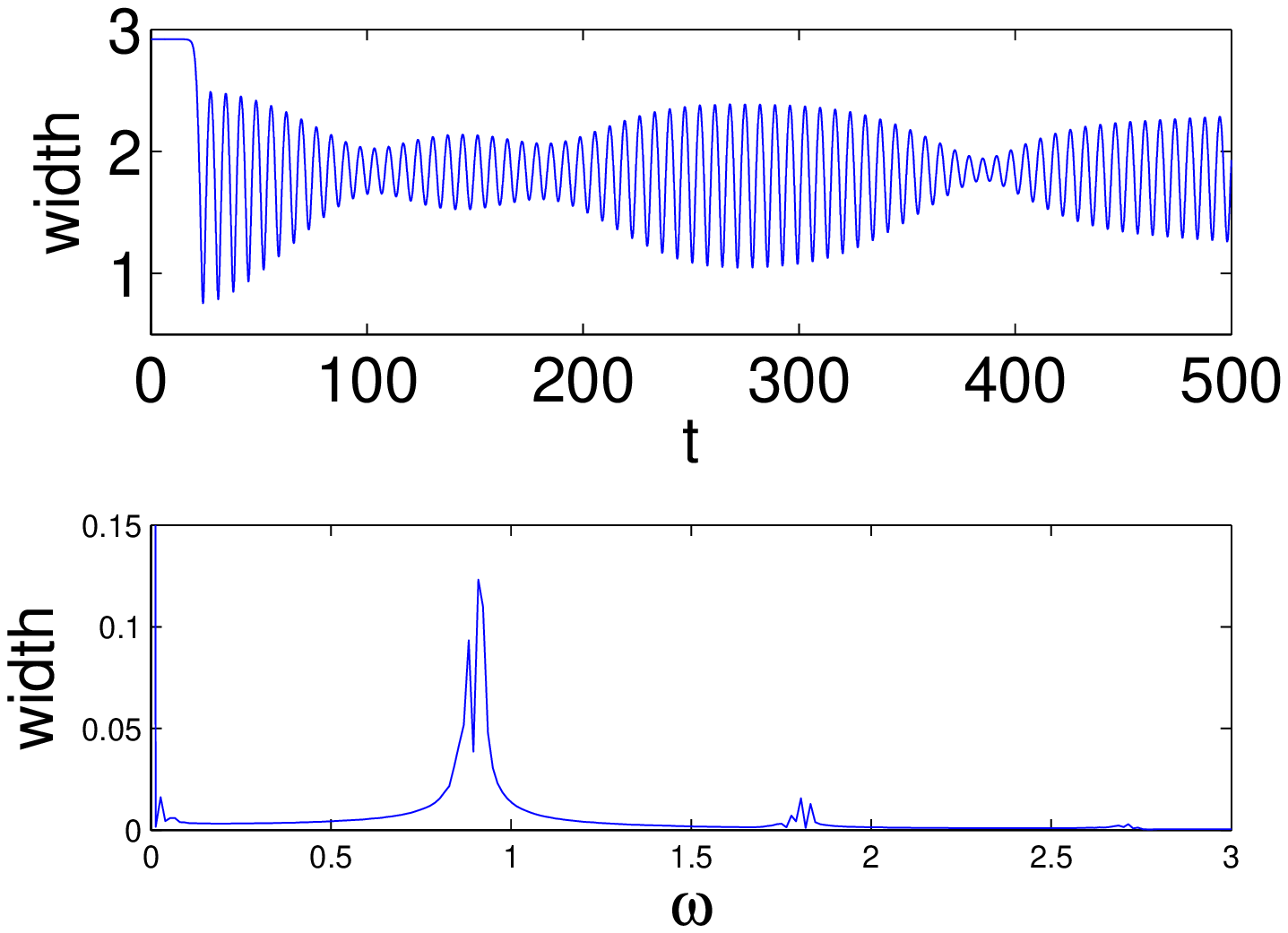} &
\includegraphics[width=55mm]{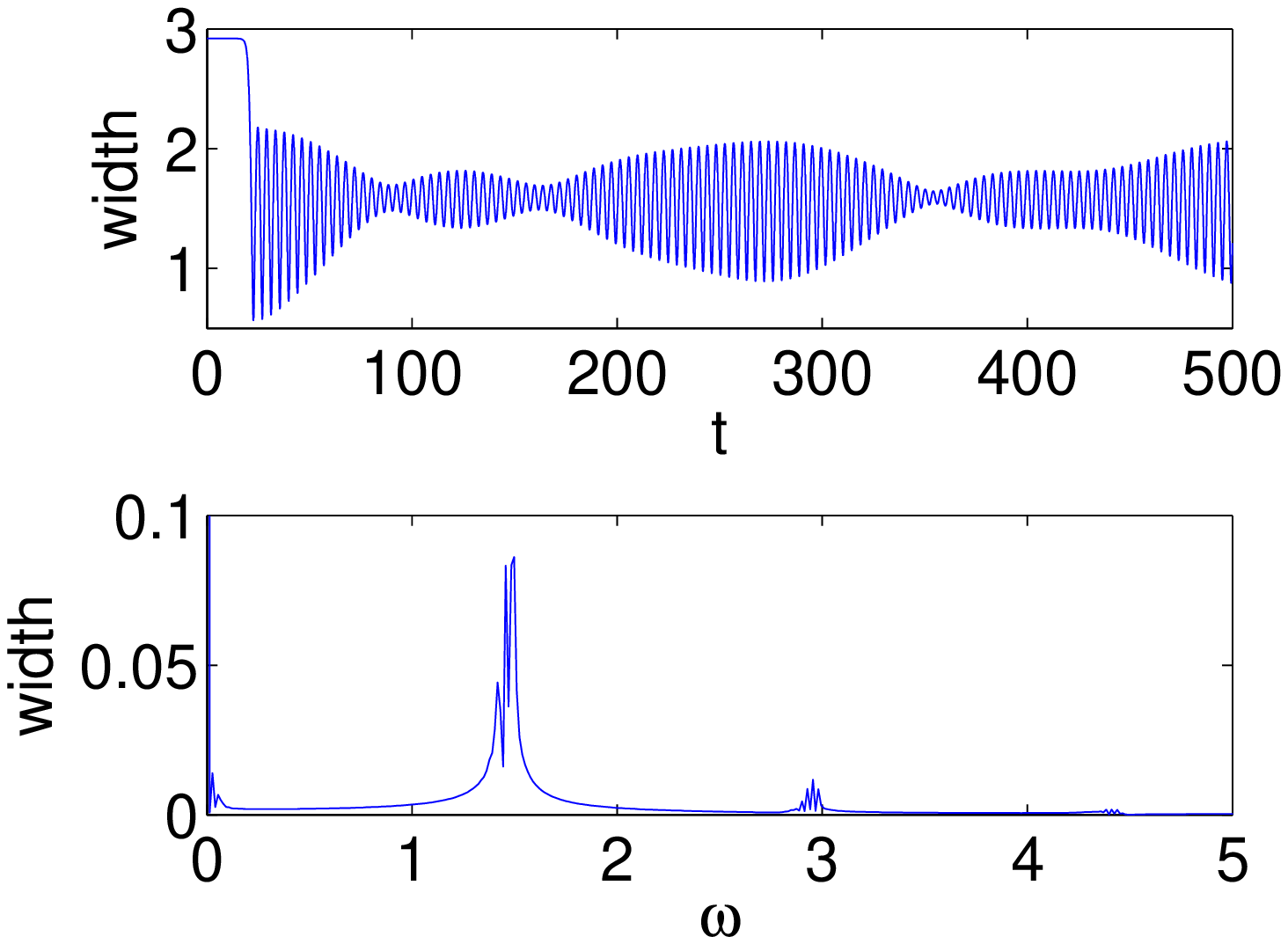} \\
\includegraphics[width=55mm]{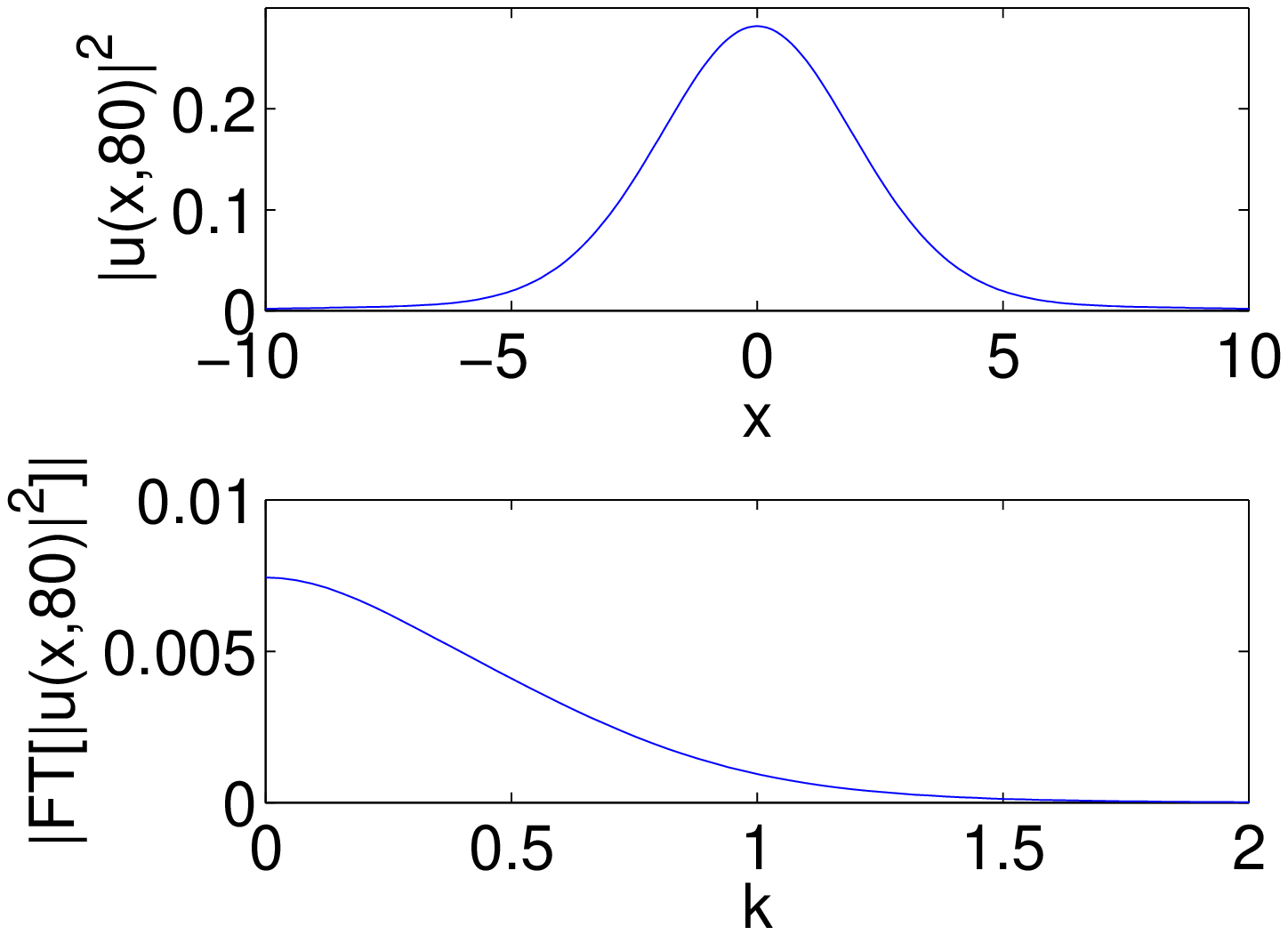} &
\includegraphics[width=55mm]{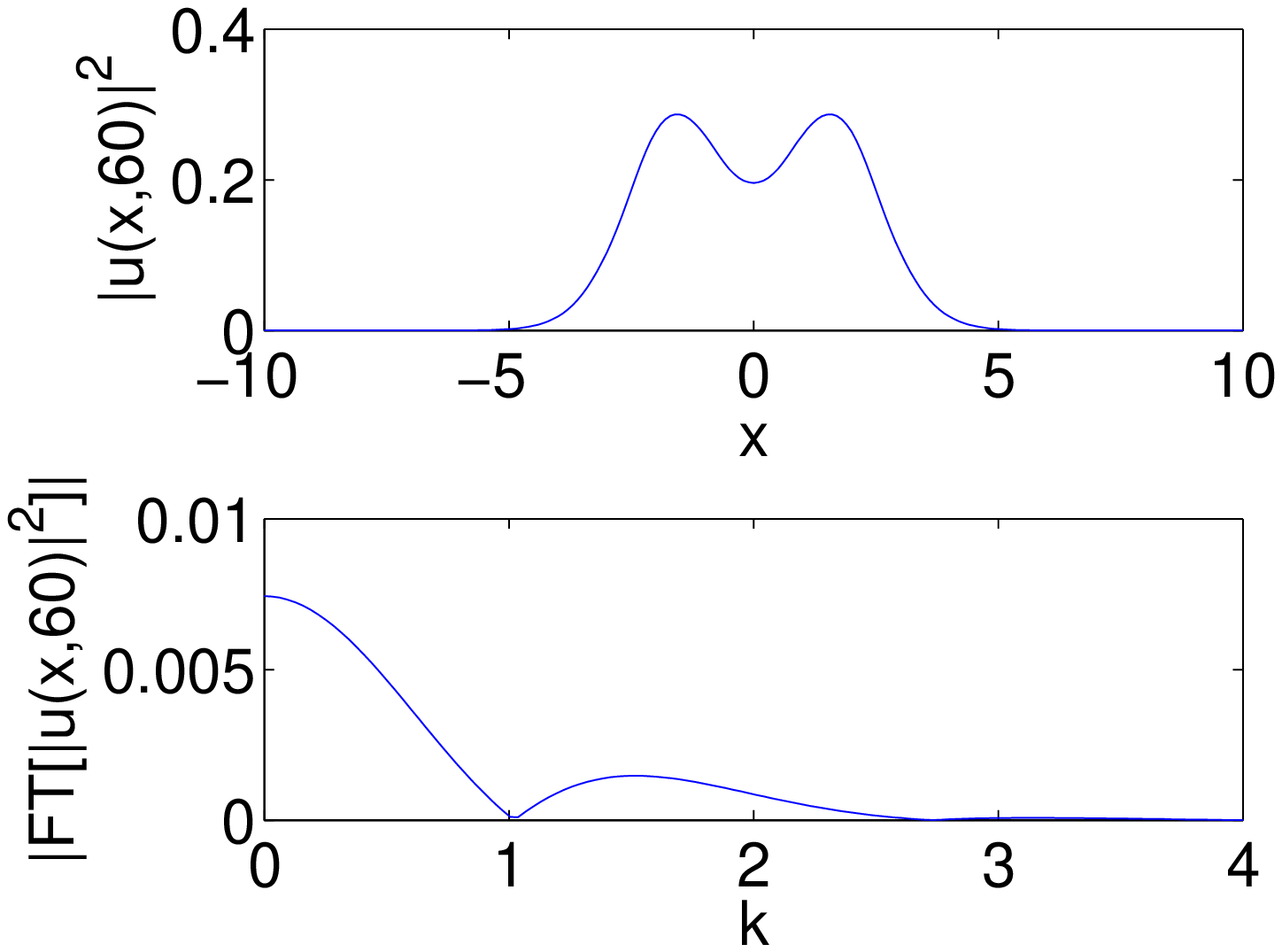} &
\includegraphics[width=55mm]{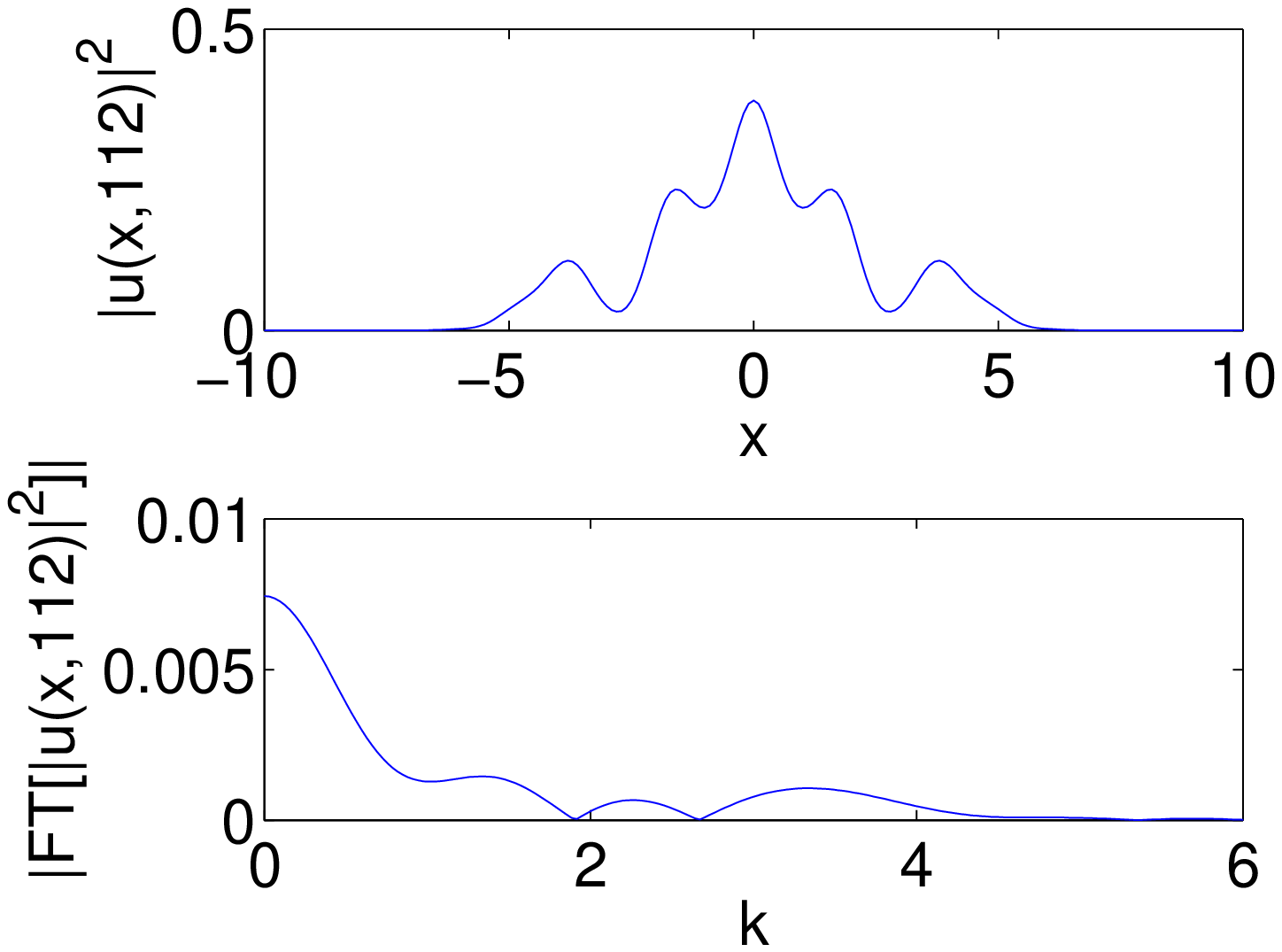}
\end{tabular}
\end{center}
\caption{Top row: The full solution intensity space-time contour
plots for $\Omega_1=0.02$, $\Omega_2=0.1$ (left), $\Omega_2=0.3$ (middle), and
$\Omega_2=0.5$ (right). Middle two rows:
The corresponding widths as a function of
time (upper panels) and frequencies (lower panels).
Bottom two rows: Sample spatial profiles (upper panels) and
corresponding  Fourier profiles (lower panels).}
\label{kfig4}
\end{figure}

Finally, we consider the case where the strength of the
harmonic trap
is increased for the same type of solutions as in Numerical Experiment III.
Here, the intensity of the solution is increased as the domain is
contracting and, hence we can simply allow the dynamics to evolve for
different values of $\Omega_2>\Omega_1$. What we observe suggests
a similar scenario of ``stability'' versus ``instability'' as in case III.
In particular, for small increase of $\Omega$ (left panels of Fig.
\ref{kfig4}), the dynamics of
the spatial profile remains unimodal (and its Fourier transform
near-Gaussian) and the beam width oscillates in
a near-periodic fashion. On the other hand, for larger trapping
strengths (middle and right panels of Fig. \ref{kfig4}), the
intensity profile involves an increasing number of modes
(and its Fourier transform accordingly has multiple nodes),
and the oscillation of the beam width becomes increasingly irregular,
involving a much wider band of frequencies. Notice that in this
case, similar to case II, no injection of matter is necessary:
the instability emerges due to the increase in the intensity,
necessitated by the conservation of the $L^2$ norm.

\section{Conclusions}

In this work, we have explored how the use of the 
domain size variations and their effects on instabilities
can be introduced in a conservative
wave setting, by examining the
prototypical case  of the one-dimensional nonlinear Schr{\"o}dinger equation.
We illustrated that this generic model has a modulational instability that
can play a role similar to the Turing instability of reaction-diffusion
systems. In particular, we demonstrated that one can take advantage of domain width variability
to controllably ``excite'' the instability and lead to the formation
of soliton train patterns in the dispersive system. We examined
four different manifestations of this feature. Two of them were relevant
to the uniform, untrapped system and two were associated with the
setting of an harmonic
trap, relevant to Bose-Einstein condensates. In the arguably most standard illustration of the
relevant phenomenology (in comparison with its dissipative analog),
expansion of a domain for a uniform, untrapped solution led to the
realization of a ``soliton sprinkler'', emitting more solitons
as the domain grew wider. On the other hand, somewhat counter-intuitively,
we showed that the instability can even emerge for untrapped systems
in contracting domains, provided that the ``mass'' ($L^2$ norm) of the
profile is preserved. We also demonstrated how the relevant stability
or instability notions can be re-interpreted in the more realistic,
yet not spatially uniform, context of harmonically
confined condensates. 
In particular, in the ``stable'' case, the distribution remained
unimodal featuring near-periodic oscillations of its width, while
in the ``unstable'' case, multiple humps emerged in the dynamics of
the intensity and complex oscillations in its moments. This was
illustrated both for the expanding domain case of decreasing the
parabolic trap strength, as well as for the contracting domain
case of increasing it.

It would certainly be relevant to implement similar ideas
in higher-dimensional settings, where there also
exists the additional complication of focusing and wave collapse \cite{sulem}.
It would be especially interesting to elucidate the interplay of
the different mechanisms, with
the modulational instability inducing the emergence of localized
waveforms, which, in turn, may be subject to collapse.
It may also be intriguing to implement such ideas in discrete systems,
given their recent extensive experimental relevance \cite{augusto,dnc0}
and some of the unique features that they possess such as
the instability being accessible even for defocusing nonlinearities.
Such studies are currently in progress and will be reported in
future publications.

{\bf Acknowledgements.}
P.G.K. gratefully acknowledges the support by NSF through the
grants DMS-0204585, DMS-CAREER, DMS-0505663 and DMS-0619492.
Work at Los Alamos is supported by the US DoE.

\end{document}